\documentclass{optica-article}

\journal{opticajournal} 

\articletype{Research Article}

\usepackage{lineno}

\begin{document}

\title{Efficient Photonic Integration of Diamond Color Centers and Thin-Film Lithium Niobate}
\author{Daniel Riedel,\authormark{1,2,\textdagger} Hope Lee,\authormark{1,\textdagger} Jason F. Herrmann,\authormark{1,\textdagger} Jakob Grzesik,\authormark{1,\textdagger} Vahid Ansari,\authormark{1} Jean-Michel Borit,\authormark{1} Hubert S. Stokowski,\authormark{1}, Shahriar Aghaeimeibodi,\authormark{1,3} Haiyu Lu,\authormark{1} Patrick J. McQuade, \authormark{1} Nick A. Melosh,\authormark{1} Zhi-Xun Shen,\authormark{1} Amir H. Safavi-Naeini,\authormark{1} Jelena Vučković,\authormark{1,*}}

\address{
\authormark{1}Edward L. Ginzton Lab, Stanford University, 348 Via Pueblo Mall, Stanford, CA, USA 94305\\
\authormark{2} Present address: AWS Center for Quantum Networking, Boston,
Massachusetts, USA\\
\authormark{3} Present address: AWS Center for Quantum Computing, San Francisco, California, USA\\
\authormark{\textdagger} These authors contributed equally to this work.\\
\email{\authormark{*} jela@stanford.edu}
}


\begin{abstract*}
On-chip photonic quantum circuits with integrated quantum memories have the potential to radically progress hardware for quantum information processing. In particular, negatively charged group-IV color centers in diamond are promising candidates for quantum memories, as they combine long storage times with excellent optical emission properties and an optically-addressable spin state. However, as a material, diamond lacks many functionalities needed to realize scalable quantum systems. Thin-film lithium niobate (TFLN), in contrast, offers a number of useful photonic nonlinearities, including the electro-optic effect, piezoelectricity, and capabilities for periodically-poled quasi-phase matching. Here, we present highly efficient heterogeneous integration of diamond nanobeams containing negatively charged silicon-vacancy (SiV) centers with TFLN waveguides. We observe greater than 90\% transmission efficiency between the diamond nanobeam and TFLN waveguide on average across multiple measurements. By comparing saturation signal levels between confocal and integrated collection, we determine a $10$-fold increase in photon counts channeled into TFLN waveguides versus that into out-of-plane collection channels. Our results constitute a key step for creating scalable integrated quantum photonic circuits that leverage the advantages of both diamond and TFLN materials.
\end{abstract*}

\section{Introduction}\label{sec1}
Optically addressable solid-state spin qubits are promising building blocks for scalable quantum networking applications \cite{Awschalom2018,Atature2018}. Among these, diamond color center defects are leading candidates for advancing quantum networks. Long-lived $^{13}$C nuclear spins in diamond can be harnessed as quantum memories \cite{Abobeih2022,Bradley2019}, while nitrogen-vacancy centers have been used to demonstrate a multi-node quantum network \cite{Pompili2021} and quantum teleportation between non-neighboring nodes \cite{Hermans2022}. Further progress hinges on the efficient integration of color centers into photonic structures to enhance the spin-photon interface without introducing excessive noise \cite{Faraon2012,Riedel2017,Orphal2023}.
 
Negatively charged group-IV color centers benefit from structural inversion-symmetry, which, to first-order, shields the local spin structure from nearby electric field noise introduced during nanofabrication\cite{Evans2016, Rugar2020b}. In addition, this family of color centers consistently exhibits high Debye-Waller factors, leading to strong emission into the zero phonon line (ZPL) \cite{Thiering2018}. These features, in combination with the efficient coupling to nanophotonic cavities, have enabled the demonstration of memory-enhanced quantum communication \cite{Bhaskar2020} and the generation of streams of indistinguishable photons \cite{Knall2022}, rendering SiV centers a prime candidate for quantum networking applications.
 
\begin{figure}[th]
\centering
\includegraphics[width=\textwidth]{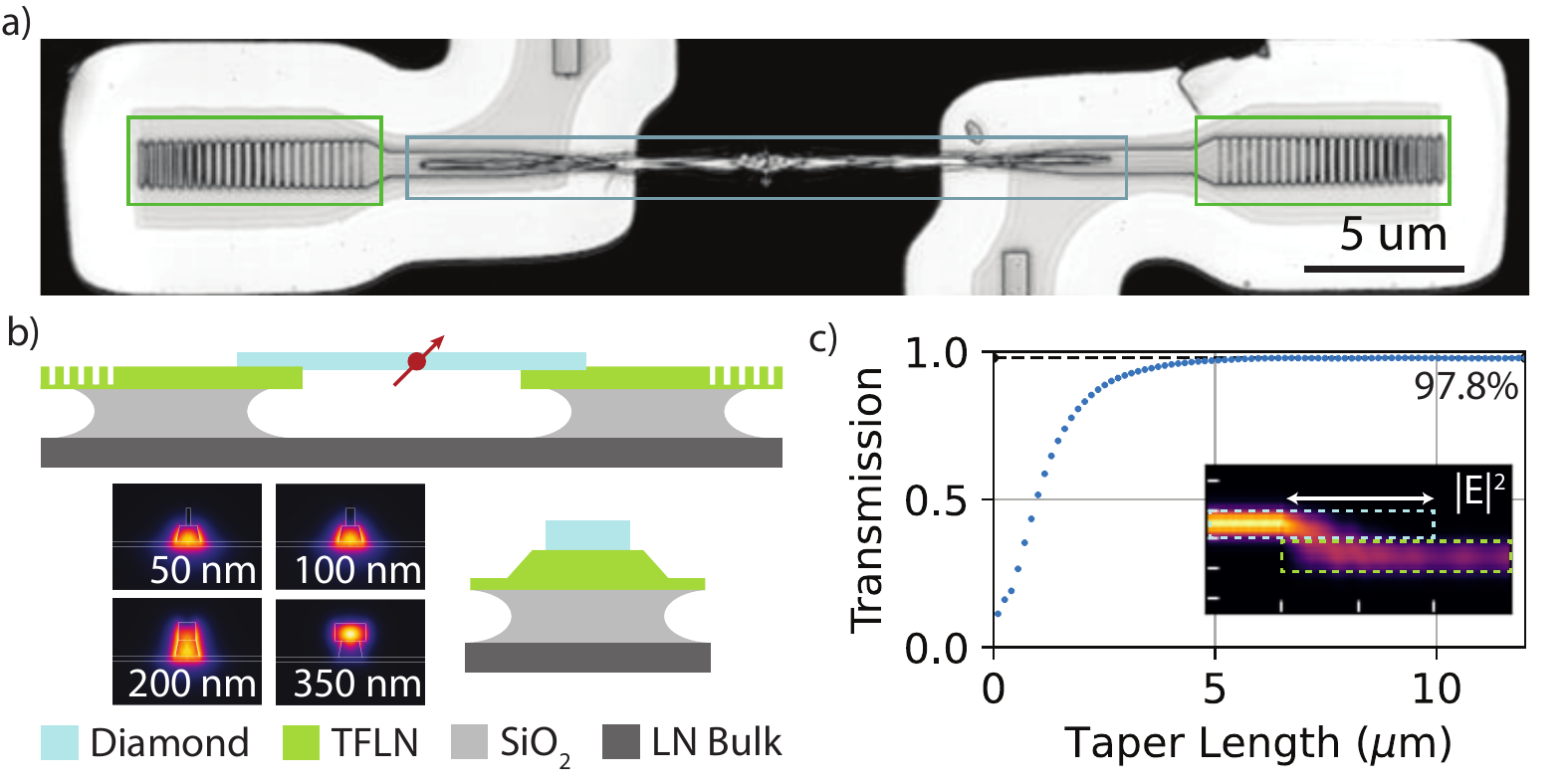}
\caption{
Structure for adiabatic transfer of light from diamond to thin-film lithium niobate 
\textbf{(a)}  Reflectance image of the integrated device with a $408$ nm laser scanning system. The green boxes indicate the TFLN grating couplers, and the blue box indicates the diamond nanobeam. 
\textbf{(b)} Schematics of the device: (top) side-profile, (bot) cross-section of the inverse taper region. A diamond nanobeam rests on thin-film lithium niobate on insulator, with a trench undercut. The insets show COMSOL\texttrademark\ simulations of the fundamental TE mode of the hybridized waveguide, demonstrating adiabatic mode transfer between the diamond and TFLN. The width of the diamond waveguide is given in each inset, with the TFLN waveguide top width fixed at $200$ nm. Lighter color indicates greater intensity.
\textbf{(c)} Lumerical FDTD simulations of transmitted power between the diamond and TFLN with swept taper lengths. The expected taper transmission efficiency for the $10$ $\mu$m designed taper length is $\sim98\%$. The inset shows the horizontal cross section of the simulated electric field, with the blue and green dashed lines acting as guides to the viewer of the approximate locations of the diamond and TFLN waveguides. Lighter color indicates greater field intensity. The horizontal scale bar indicates a $10$ $\mu$m overlap in the tapers, while the vertical tick marks are spaced by $200$ nm.} 
\label{fig:fig1}
\end{figure}

However, scalable, monolithic fabrication of diamond photonic integrated circuits (PICs) remains challenging, and complex diamond PICs are precluded by the absence of second-order nonlinearities. To overcome these challenges, heterogeneous integration with mature nanophotonic material platforms is a promising approach \cite{Wan2020,Xuan2016,Lu2019a,Elshaari2020,Kim2020,Lake2016,Wilson2020,Jung2021,Guidry2020,Wang2021,Wu2018,Zhu2021}. Thin-film lithium niobate (TFLN) offers many advantages over alternative photonic materials via a large $\chi^{(2)}$ nonlinearity. This enables many applications, including strong electro-optic modulation and frequency conversion \cite{wang2018nanophotonic,zhang2021integrated,mckenna2020cryogenic,holzgrafe2020cavity}, piezoelectric transduction \cite{jiang2019lithium,jiang2022optically}, and periodic poling for quasi-phase matching and nonlinear frequency conversion \cite{mckenna2022ultra,park2022high}. 

Here, we demonstrate efficient heterogeneous integration of a diamond nanobeam featuring incorporated SiV color centers with a TFLN platform using a mechanical pick-and-place approach\cite{Mouradian2017,Zhu2022,Chanana2022,Wan2020}. By precisely placing double-tapered diamond nanobeams, we demonstrate the bridging of a gapped TFLN waveguide with a diamond-to-LN transmission efficiency of $92\pm11\%$ per facet at $737$ nm, corresponding to the SiV ZPL wavelength, averaged across multiple measurements. We find a nearly $2.5$-times improvement in ZPL photon extraction via integrated TFLN collection channels compared to out-of-plane collection from the same device. By taking into consideration grating coupler efficiencies, we can infer an approximately $10$-fold improvement in photon channeling into the TFLN waveguide compared to confocal collection channels. Our results demonstrate a crucial step towards the incorporation of high-quality diamond spin qubits into a scalable nonlinear photonics platform.

\section{Results}\label{sec2}

\begin{figure}[t]
\centering
\includegraphics[]{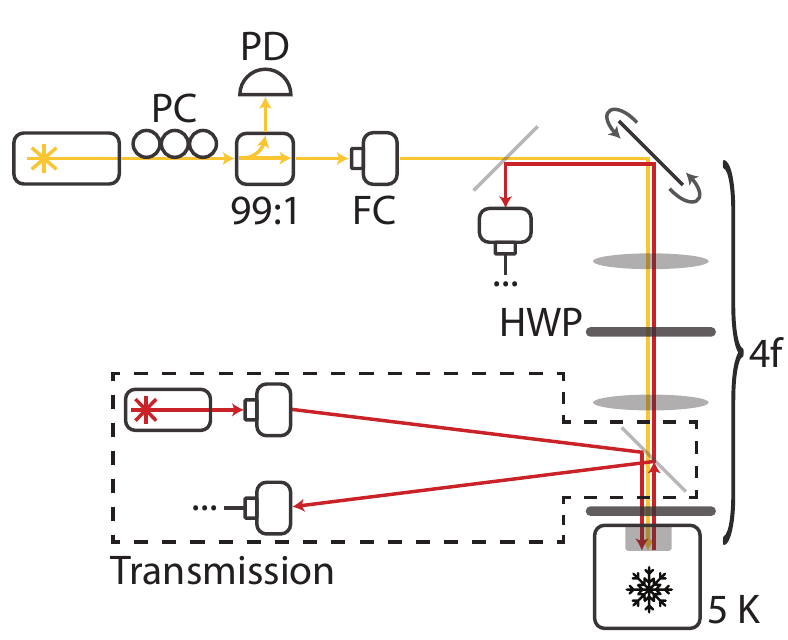}
\caption{Schematic of measurement setup for optical characterizations. ``PC'' (polarization controller), ``PD'' (photodiode), ``$99:1$'' ($99$\%$:1$\% beamsplitter), ``$4$f'' ($4$-f imaging path, lengths not to scale), ``HWP'' (half-wave plate), ``FC'' (fiber-to-free space coupler), ``$5$ K'' ($5$ Kelvin). The dashed box indicates the removable dichroic beamsplitter and transmission path. We achieve spatial separation of the two transmission collection spots via a D-shaped mirror.}
\label{fig:fig2a}
\end{figure}

\subsection{Device Fabrication and Transfer}\label{subsec1}
Our device consists of a diamond double-tapered nanobeam with incorporated SiV centers bridging a gapped, undercut TFLN waveguide, imaged in Fig.\ref{fig:fig1}(a). The SiV color centers are generated during a chemical vapor deposition (CVD) overgrowth process on a electronic grade bulk single-crystalline diamond. We fabricate the diamond nanobeams using the well-established quasi-isotropic etching technique following the procedures outlined in \cite{Xie2018,Khanaliloo2015,Khanaliloo2015b,Mouradian2017,Rugar2020b} (see supplemental section S2). Our diamond devices consist of a $10$ $\mu$m long rectangular nanobeam with $10$ $\mu$m tapers at each end, for a total length of $30$ $\mu$m. The nanobeam has a target thickness of $\sim 200$ nm and its width is tapered from $\sim 350$ nm to $\sim 50$ nm at the taper end. The device is anchored to the bulk substrate via a thin tether, around which the nanobeam is widened slightly to reduce scattering effects of the tether on optical transmission.

The TFLN waveguides are fabricated following the techniques demonstrated in \cite{mckenna2020cryogenic}. We first pattern a negative resist mask atop a lithium niobate-on-insulator (LNOI) substrate using electron beam lithography. We then employ Ar ion milling to etch the waveguides. A second photolithography mask and ion mill step is used for additional removal of the TFLN slab to expose the buried oxide layer of the LNOI. An additional acid cleaning procedure partially etches this oxide, thereby undercutting the waveguide sockets. The completed TFLN waveguides are approximately $190$ nm thick atop an approximately $60$ nm thick slab and $\sim1$ $\mu$m wide, which assists in the placement of the diamond nanobeam by allowing for a larger margin of alignment error (Fig.\ref{fig:fig1}(b)). The waveguides adiabatically taper down to $\sim100$ nm over a length of $5$ $\mu$m. A $15$ $\mu$m gap is left between the two waveguides to serve as a ``socket'' for a diamond nanobeam. Each waveguide ends in a grating optimized to couple $737$ nm light to and from the device (see supplemental section S7). 

Following fabrication, the diamond nanobeam is transferred via pick-and-place to the TFLN socket. The mechanical transfer process is carried out using a home-built micro-manipulation setup equipped with confocal imaging capabilities and dual tungsten ``cat-whisker'' needles, each with a $\sim70$ nm tip radius. To assist with the mechanical break-off process, the holding tethers are partially cut via focused ion beam (FIB) etching. During transfer, precise orientation of the nanobeam can be controlled with the dual needles, while the LN chip is mounted on a combination of translation and rotation stages. After careful positioning, strong adhesion between the diamond nanobeam and the TFLN via Van der Waals forces allows us to remove the attached needle by pulling it down into the etched trench between the TFLN waveguides. Ultimately, the diamond nanobeam tapers are positioned inversely to the LN ``socket'' tapers, enabling high-efficiency adiabatic transfer of light between the TFLN waveguide and the diamond nanobeam, as shown in Fig.\ref{fig:fig1}(b). 

\begin{figure}[t]
\centering
\includegraphics[]{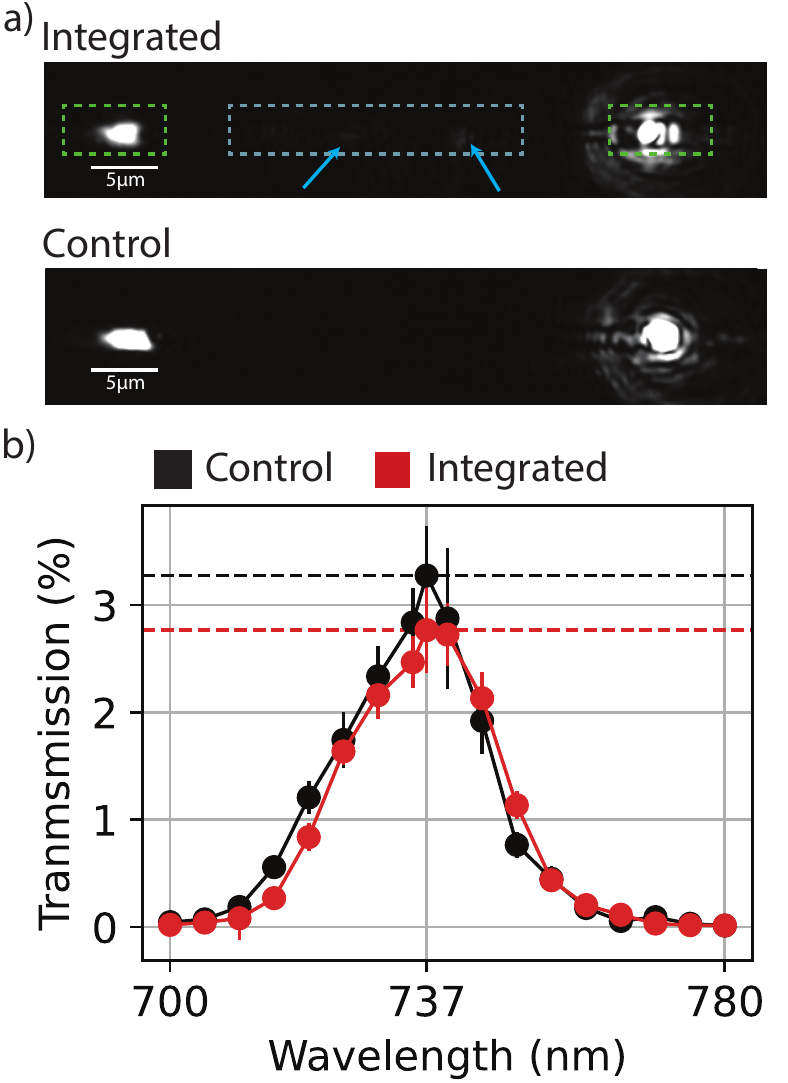}
\caption{Transmission characterization through the integrated device.
\textbf{(a)} (top) White light image of efficient transmission of light through the integrated device, indicating high taper efficiency at the diamond/TFLN interface. Minimal scattering (indicated by arrows) at the contact points and along the device further demonstrates efficient abiabatic transfer of light. The diamond nanobeam and grating coupler locations are indicated by the blue and green dashed boxes, respectively. (bot) White light image for comparison of the fully-connected TFLN control device. Images taken at $20$ ms camera exposure.
\textbf{(b)} Average transmission through the integrated diamond-TFLN device (red) and a fully-connected TFLN ``control'' device (black). Error bars depict the standard deviation by averaging four measurements over two separate cooldowns of the devices. The dashed black and red lines indicate the transmission at $737$ nm for the control and integrated devices respectively.}
\label{fig:fig2}
\end{figure}

\subsection{Optical Characterization of Devices}\label{subsec2}
We characterize our device in a closed-cycle Montana cryostat at a temperature of $\sim$5 K utilizing a home-built confocal microscopy setup consisting of two distinct access arms. We refer to these collection arms as the confocal and transmission paths. The transmission path can be removed from the apparatus by withdrawing the dichroic just before the cryostat. The full measurement apparatus is schematically depicted in Fig.\ref{fig:fig2a}.

\subsubsection{Transmission}\label{subsec3}
As a first step, we measure transmission through our device, using the TFLN grating couplers to characterize the single-taper transmission efficiency between the diamond and LN. Using the transmission measurement path, we send a narrow-band laser source through one grating coupler and measure the output power through the other. The input laser power is calibrated via a fiber beamsplitter and photodiode at the input. The output transmitted light is coupled into a single mode fiber and routed to a second photodiode. We sweep the Ti:Sapphire laser excitation from $700$ nm to $780$ nm in intervals of $5$ nm, with a separate measurement performed at $737$ nm, specifically, as demonstrated in Fig.\ref{fig:fig2}(b). Direct transmission through the device at $737$ nm is determined to be an uncorrected $2.8\pm0.4$\% from single-mode fiber to single-mode fiber, averaged over four distinct measurements. We then shift the stage to a ``control'' device, consisting of a fully-connected (i.e., no taper or gap) TFLN waveguide with nominally identical grating couplers, while keeping the excitation and collection alignment fixed. Transmission through the control device (similarly averaged across four distinct measurements) is $3.3\pm0.5$\%, implying a single-mode grating coupler efficiency of $17.3\pm1.2$\%, in agreement with grating coupler simulations (see supplemental section S7). Attributing all additional losses in the device to the adiabatic taper facets, we conclude that each diamond-LN taper has a transmission efficiency of $92\pm11\%$ at $737$ nm assuming equal taper and grating efficiencies (Fig.\ref{fig:fig2}(b) – see supplemental section S5 for additional details on the calibration procedure). This high efficiency is visualized by the lack of significant scattering at both taper contact points to the TFLN and along the diamond nanobeam in Fig.\ref{fig:fig2}(a). Importantly, grating couplers are very sensitive to the angle and focal plane of incident light. Therefore, crosstalk and hysteresis in our stage motion can introduce error into this efficiency calibration. However, additional repeated measurements across multiple thermal cycles yield similarly consistent values for taper efficiency. Furthermore, while this is the highest-efficiency device we measured, a second device exhibits an estimated single-taper transmission efficiency of greater than $80$\% (see supplemental section S5).

\begin{figure}[t!]
\centering
\includegraphics[width=\textwidth]{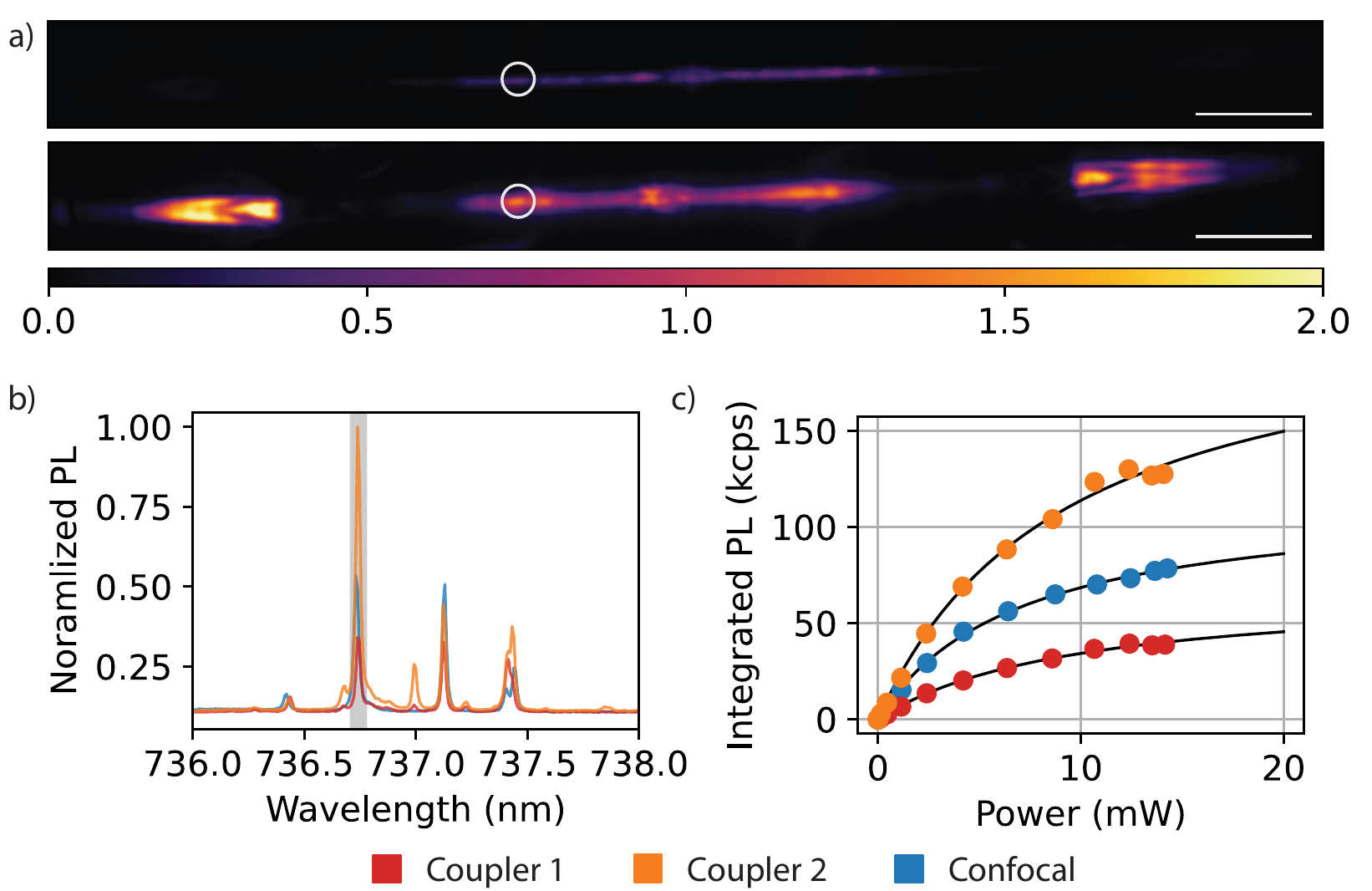}
\caption{
Color center PL characterizations. 
\textbf{(a)} PL scans, with confocal collection (top) and stationary coupler collection (bottom), reported in counts/sec. The coupler collection PL scan presented is obtained by counting photon arrivals from the couplers in parallel, with each avalanche photodiode collecting emission from one grating coupler. Over an integration period, these simultaneous counts are summed to yield the total photons collected from both grating couplers. Excitation for both were provided confocally, and scanning was performed utilizing a galvanometer mirror and 4f setup. The white circle in each scan indicates the location of confocal excitation for panels (b) and (c). Scale bars indicate 5$\mu$m. The colorbar is expressed in units of $1e6$ counts per second (cps).
\textbf{(b)} Normalized PL spectrum, overlaid with confocal and both coupler collected spectra. The grey region indicates the post-processed 'filtered' wavelength range, determined by fitting the spectra to multiple Lorentzian peaks and identifying the FWHM of the selected emission peak for a single emitter.
\textbf{(c)} Saturation curves for each collection configuration. CPS for each power and configuration was determined by integrating the number of counts collected in a spectra within the FWHM ranges identified by Lorentzian fit. PL spectra were each integrated for $2.5$ s. The black trend lines represent a standard power saturation fit. Error bars for each data point was calculated using Poissonian photon statistics but are smaller than data markers. 
} 
\label{fig:fig3}
\end{figure}

\begin{figure}[t!]
\centering
\includegraphics[width=\textwidth]{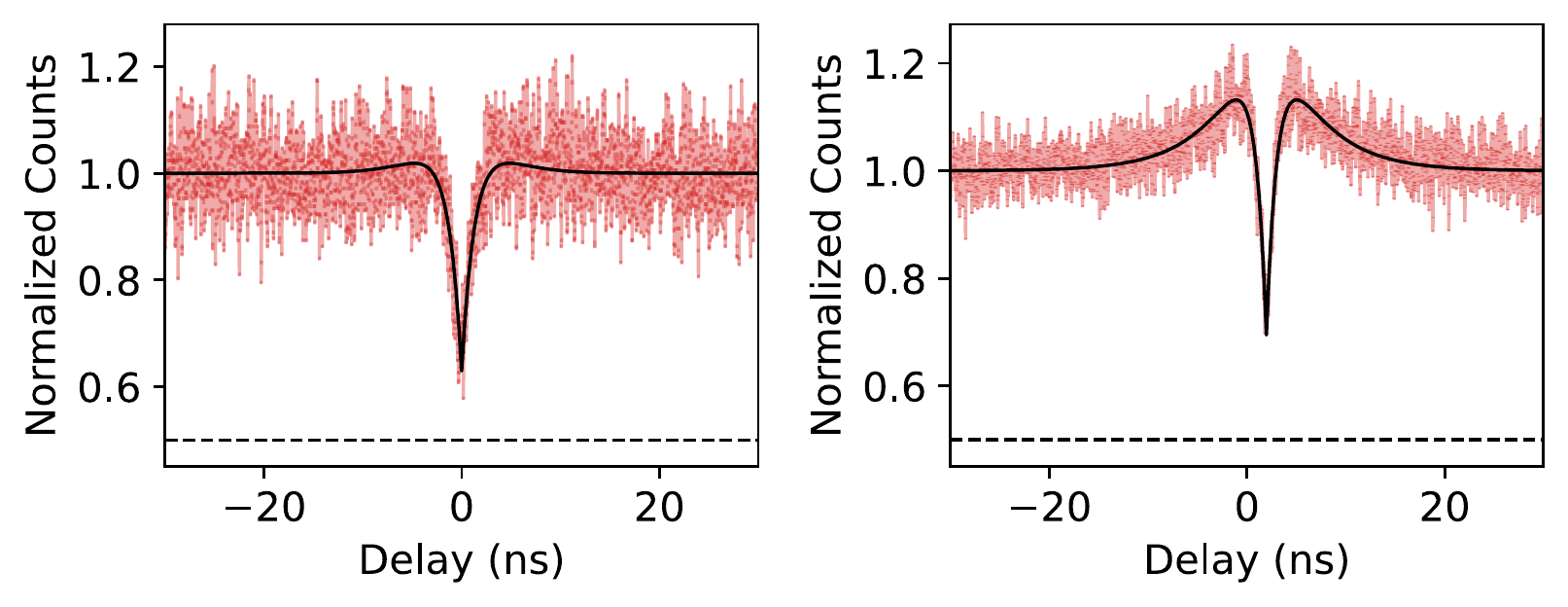}
\caption{ Photon autocorrelation measurements 
A confocally collected (coupler collected) autocorrelation measurement with $g^{2}(0)=0.629 \pm 0.006$ ($g^{2}(0)=0.678 \pm 0.004$) on the left (right). The data was fit using a bunched $g^{2}(\tau)$ function form, with counts far from the delay dip normalized to $1$. The dashed line indicates the $0.5$ threshold for single photon emitter characteristics. Error bars were calculated assuming Poissonian photon statistics. 
} 
\label{fig:fig4}
\end{figure}

\subsubsection{Optical Addressing of SiV}\label{subsec4}
We next perform photoluminescence (PL) measurements on SiV color centers embedded in the diamond waveguide. There are two measurement configurations: one with the excitation and collection co-localized via the confocal path (C/C), and one with confocal excitation but collection through the grating couplers and transmission path (C/CP). Using a homebuilt confocal scanning microscope, we excite color centers off-resonantly at $710$ nm and collect the ZPL emission at $737$ nm. The resulting PL maps and spectra are presented in Fig.\ref{fig:fig3}(a). After locating isolated bright spots in PL, we sweep the optical excitation power and record the PL spectra, comparing the efficiency of SiV emission collected confocally to that collected through the integrated, nanobeam channel. We toggle between confocal and grating coupler collection paths by removing the transmission path dichroic and adjusting only the half-wave plate just before the cryostat. We are careful to make minimal extraneous optical adjustments in order to maintain the validity of any comparisons of the collected signal.

From the PL spectra, we observe a small ensemble of emitters arising from the homogeneous distribution of SiVs throughout the diamond nanobeam. Therefore, for our specific device, we are unable to isolate a single emitter through off-resonant confocal excitation. To quantify photon extraction efficiencies, we thus perform saturation measurements first with the C/CP configuration, then with the C/C configuration, and integrate within $6\times$ the FWHM of identified Lorentzian peaks. As further post filtering, we selectively integrate and fit to the saturation model a representative transition for an SiV center, demarcated by gray in Fig.\ref{fig:fig3}(b). We note that we selected a transition which couples favorably to the diamond nanobeam. Errors for photon counts are given by assuming Poissonian photon statistics. We determine a $6.9\pm0.2$ mW ($9.8\pm0.7$ mW, $9.3\pm0.7$ mW) saturation power and $116\pm2$ kcps ($68\pm2$ kcps, $219\pm8$ kcps) saturation count rate for C/C configuration (C/CP$_1$, C/CP$_2$). We attribute the slight discrepancy between C/C and C/CP saturation powers to polarization adjustments made between the configurations in order to optimize power delivery to the color center with the removal of the second dichroic mirror \cite{Mouradian2015}. The difference in signal between the two CP configurations can be attributed to preferential emission of the SiV in one direction of the nanobeam due to either angular dipole orientation or positioning in the nanobeam (see supplemental section S6). Most notably, we observe a nearly $2.5$-times increase in color center photon extraction through the diamond nanobeam and TFLN waveguide grating couplers compared to the confocal channel for collection through CP, indicating extremely high-efficiency diamond-TFLN coupling. Correcting for the grating coupler efficiency $<25\%$, we infer that channeling of SiV photons into the TFLN waveguide is improved by more than $10$-fold when compared to out-of-plane confocal collection through a high-NA objective.  

As further evidence of emitter co-localization, we measure $g^{(2)}(\tau)$ auto-correlation of its emission in the C/C configuration using a $50:50$ fiber beamsplitter. We  isolate a single emission line with a series of filters, and observe anti-bunching of $g^{(2)}(0)=0.629 \pm 0.006$, indicating emission from an ensemble of emitters within our filtered wavelength region, as shown in Fig.\ref{fig:fig4}(a). We then repeat autocorrelation measurements in the C/CP configuration, utilizing our device as an integrated beamsplitter to correlate emission between the two grating couplers. We measure a similar $g^{(2)}(0)=0.678 \pm 0.004$ with this approach, depicted in Fig.\ref{fig:fig4}(b). The power dependence of the auto-correlation is presented in supplemental section S8.

Lastly, we consider the robustness of the integration to environmental factors. It should be noted that this particular device survived multiple partial and complete thermal cycles. Furthermore, during one such event, the nanobeam fully detached from the TFLN waveguide socket and the pick-and-place apparatus was used to re-position the nanobeam. In all cases, we observed similarly high transmission efficiency. Additional controlled thermal cycling and targeted testing will be necessary to formally characterize this robustness, but repeated transmission measurements suggest a lack of deterioration in device performance.

\section{Discussion}\label{sec4}
Overall, we have demonstrated the photonic integration of diamond color centers with thin-film lithium niobate. Our integration yields very high coupling efficiency of $91.9\pm11\%$ per-facet between the diamond and the TFLN. This translates into a nearly $2.5$-times increase in off-chip photon collection through integrated channels, and a $10$-fold enhancement in photon channeling into the LN waveguides versus out-of-plane. These metrics enable us to combine the promise of diamond spin qubits with the advantageous optical nonlinearities native to lithium niobate. 

Furthermore, the emission count rate is potentially limited by random placement and non-ideal angular orientation of the SiV inside the nanobeam. The homogeneous distribution of color centers within the nanobeam also complicates the probing of single emitter properties in our device. These challenges can be overcome by delta doping and targeted implantation techniques, such as masked or focused ion beam implantation \cite{Nguyen2019a, Rugar2020, Schroder2017, Titze2022}. Furthermore, photon extraction from the emitters into the TFLN waveguide can be enhanced by using resonant photonic structures \cite{Rugar2021a,Nguyen2019b}.

A major advantage of selective integration is the possibility for large-scale pre-characterization of diamond nanobeams and emitters, enabling a more deterministic device yield \cite{Sutula2022}. The transfer process could potentially be fully automated using a combination of machine vision and \textit{in situ} transmission measurements.

Our device paves the way for high-efficiency emitter collection on TFLN, enabling single-photon experiments that leverage TFLN's unique properties. We aim to utilize optical nonlinearities in LN for modulating color center emission, thereby demonstrate multi-emitter interference and multiplexing on chip \cite{Narita2022}. We further aspire to add additional on-chip capabilities, such as on-chip filters \cite{Elshaari2017} and detectors \cite{Reithmaier2015,Najafi2015}. On-chip photon routing would enable both the entanglement between several integrated quantum memories and coupling to the same external fiber channel. In combination with recently demonstrated high fiber packaging efficiencies \cite{zeng2023fiberintegration}, this would enable integrating a large number of quantum memories on the same chip.
We also aim to leverage periodic poling in LN to achieve quantum frequency conversion of emitters to the infrared band for integration with the telecom band. These future directions would pave the way towards on-chip heterogeneous platforms linking the excellent optical and coherence properties of solid state spin qubits with the potential of a mature nonlinear photonics platform.

\backmatter

\section*{Supplementary information}
Supplementary material provided.

\section*{Acknowledgments}
This material is based upon work supported by the U.S. Department of Energy Office of Science National Quantum Information Science Research Centers as part of the Q-NEXT center. H. Lee acknowledges support from the Stanford Graduate Fellowship. JFH acknowledges support from the NSF GRFP under grant no. DGE-1656518. JG acknowledges support from the Hertz Foundation Graduate Fellowship. SA and VA acknowledge support from the Bloch postdoctoral fellowship in quantum science and engineering from Stanford Quantum Fundamentals, Architecture and Machines initiative (Q-FARM). DR acknowledges support from the Swiss National Science Foundation (project no. P400P2\_194424). DR and SA contributed to this work prior to joining AWS. HSS acknowledges support from the Urbanek Family Fellowship. H. Lu and ZXS were supported by the DOE Office of Basic Energy Sciences, Division of Materials Science and Engineering. The authors thank Patrick McQuade, and Alison Rugar for assistance with diamond sample growth. The authors thank Daniil Lukin, Samuel Gyger, Eric Irving Rosenthal, and Luke Qi for helpful discussions regarding sample measurements and data analysis. Work was performed in part in the nano@Stanford labs, which are supported by the National Science Foundation as part of the National Nanotechnology Coordinated Infrastructure under award ECCS-2026822. Part of this work was performed at the Stanford Nano Shared Facilities (SNSF), supported by the National Science Foundation under award ECCS-2026822.

Note. During the writing of this paper we
learned of work using similar techniques to calibrate for the transmission efficiencies of adiabatic tapers in a different material platform \cite{larocque2023tunable}.

\section*{Declarations}
The authors declare no competing interests.
\section*{Data Availability}
All data is available from the authors upon reasonable request.

\section*{Authors' Contributions}
H. Lee, JG, and DR contributed to the diamond nanobeam design and fabrication. H. Lu, PM, and ZXS, contributed to the diamond sample growth and material preparation. TFLN grating couplers were designed by JFH, and JFH and VA contributed to the TFLN chip fabrication. The adiabatic taper was initially designed and simulated by JMB, DR, and SA, with follow-up simulations conducted by JMB, DR, and JFH. The pick and place setup and transfer of the diamond nanobeam to TFLN was carried out by H. Lee, DR, and JG. Optical measurements of the integrated device were led by H. Lee and assisted by DR, JG, and JFH. The manuscript was prepared by H. Lee, JFH, JG, and DR. Additional support and assistance were contributed by HSS, ASN, and JV. ASN and JV oversaw the experiment and assisted in data interpretation and editing of the manuscript. 


\bibliography{LND_Bib}

\begin{thebibliography}{10}
\newcommand{\enquote}[1]{``#1''}

\bibitem{Awschalom2018}
D.~D. Awschalom, R.~Hanson, J.~Wrachtrup, and B.~B. Zhou, \enquote{{Quantum
  technologies with optically interfaced solid-state spins},}
  {\protect\JournalTitle{Nature Photonics}} \textbf{12}, 516--527 (2018).

\bibitem{Atature2018}
M.~Atat{\"{u}}re, D.~Englund, N.~Vamivakas, S.-Y. Lee, and J.~Wrachtrup,
  \enquote{{Material platforms for spin-based photonic quantum technologies},}
  {\protect\JournalTitle{Nature Reviews Materials}} \textbf{3}, 38--51 (2018).

\bibitem{Abobeih2022}
M.~H. Abobeih, Y.~Wang, J.~Randall, S.~J. Loenen, C.~E. Bradley, M.~Markham,
  D.~J. Twitchen, B.~M. Terhal, and T.~H. Taminiau, \enquote{{Fault-tolerant
  operation of a logical qubit in a diamond quantum processor},}
  {\protect\JournalTitle{Nature}} \textbf{606}, 884--889 (2022).

\bibitem{Bradley2019}
C.~E. Bradley, J.~Randall, M.~H. Abobeih, R.~C. Berrevoets, M.~J. Degen, M.~A.
  Bakker, M.~Markham, D.~J. Twitchen, and T.~H. Taminiau, \enquote{{A Ten-Qubit
  Solid-State Spin Register with Quantum Memory up to One Minute},}
  {\protect\JournalTitle{Physical Review X}} \textbf{9}, 031045 (2019).

\bibitem{Pompili2021}
M.~Pompili, S.~L.~N. Hermans, S.~Baier, H.~K.~C. Beukers, P.~C. Humphreys,
  R.~N. Schouten, R.~F.~L. Vermeulen, M.~J. Tiggelman, L.~{dos Santos Martins},
  B.~Dirkse, S.~Wehner, and R.~Hanson, \enquote{{Realization of a multinode
  quantum network of remote solid-state qubits},}
  {\protect\JournalTitle{Science}} \textbf{372}, 259--264 (2021).

\bibitem{Hermans2022}
S.~L.~N. Hermans, M.~Pompili, H.~K.~C. Beukers, S.~Baier, J.~Borregaard, and
  R.~Hanson, \enquote{{Qubit teleportation between non-neighbouring nodes in a
  quantum network},} {\protect\JournalTitle{Nature}} \textbf{605}, 663--668
  (2022).

\bibitem{Faraon2012}
A.~Faraon, C.~Santori, Z.~Huang, V.~M. Acosta, and R.~G. Beausoleil,
  \enquote{{Coupling of nitrogen-vacancy centers to photonic crystal cavities
  in monocrystalline diamond},} {\protect\JournalTitle{Physical Review
  Letters}} \textbf{109}, 33604 (2012).

\bibitem{Riedel2017}
D.~Riedel, I.~S{\"{o}}llner, B.~J. Shields, S.~Starosielec, P.~Appel, E.~Neu,
  P.~Maletinsky, and R.~J. Warburton, \enquote{{Deterministic Enhancement of
  Coherent Photon Generation from a Nitrogen-Vacancy Center in Ultrapure
  Diamond},} {\protect\JournalTitle{Physical Review X}} \textbf{7}, 031040
  (2017).

\bibitem{Orphal2023}
L.~Orphal-Kobin, K.~Unterguggenberger, T.~Pregnolato, N.~Kemf, M.~Matalla,
  R.-S. Unger, I.~Ostermay, G.~Pieplow, and T.~Schr{\"{o}}der,
  \enquote{{Optically Coherent Nitrogen-Vacancy Defect Centers in Diamond
  Nanostructures},} {\protect\JournalTitle{Physical Review X}} \textbf{13},
  011042 (2023).

\bibitem{Evans2016}
R.~E. Evans, A.~Sipahigil, D.~D. Sukachev, A.~S. Zibrov, and M.~D. Lukin,
  \enquote{{Narrow-Linewidth Homogeneous Optical Emitters in Diamond
  Nanostructures via Silicon Ion Implantation},}
  {\protect\JournalTitle{Physical Review Applied}} \textbf{5}, 044010 (2016).

\bibitem{Rugar2020b}
A.~E. Rugar, C.~Dory, S.~Aghaeimeibodi, H.~Lu, S.~Sun, S.~D. Mishra, Z.-X.
  Shen, N.~A. Melosh, and J.~Vu{\v{c}}kovi{\'{c}}, \enquote{{Narrow-Linewidth
  Tin-Vacancy Centers in a Diamond Waveguide},} {\protect\JournalTitle{ACS
  Photonics}} \textbf{7}, 2356--2361 (2020).

\bibitem{Thiering2018}
G.~Thiering and A.~Gali, \enquote{{Ab Initio Magneto-Optical Spectrum of
  Group-IV Vacancy Color Centers in Diamond},} {\protect\JournalTitle{Physical
  Review X}} \textbf{8}, 021063 (2018).

\bibitem{Bhaskar2020}
M.~K. Bhaskar, R.~Riedinger, B.~Machielse, D.~S. Levonian, C.~T. Nguyen, E.~N.
  Knall, H.~Park, D.~Englund, M.~Lon{\v{c}}ar, D.~D. Sukachev, and M.~D. Lukin,
  \enquote{{Experimental demonstration of memory-enhanced quantum
  communication},} {\protect\JournalTitle{Nature}} \textbf{580}, 60--64 (2020).

\bibitem{Knall2022}
E.~N. Knall, C.~M. Knaut, R.~Bekenstein, D.~R. Assumpcao, P.~L. Stroganov,
  W.~Gong, Y.~Q. Huan, P.-J. Stas, B.~Machielse, M.~Chalupnik, D.~Levonian,
  A.~Suleymanzade, R.~Riedinger, H.~Park, M.~Lon{\v{c}}ar, M.~K. Bhaskar, and
  M.~D. Lukin, \enquote{{Efficient Source of Shaped Single Photons Based on an
  Integrated Diamond Nanophotonic System},} {\protect\JournalTitle{Physical
  Review Letters}} \textbf{129}, 053603 (2022).

\bibitem{Wan2020}
N.~H. Wan, T.-J. Lu, K.~C. Chen, M.~P. Walsh, M.~E. Trusheim, L.~{De Santis},
  E.~A. Bersin, I.~B. Harris, S.~L. Mouradian, I.~R. Christen, E.~S. Bielejec,
  and D.~Englund, \enquote{{Large-scale integration of artificial atoms in
  hybrid photonic circuits},} {\protect\JournalTitle{Nature}} \textbf{583},
  226--231 (2020).

\bibitem{Xuan2016}
Y.~Xuan, Y.~Liu, L.~T. Varghese, A.~J. Metcalf, X.~Xue, P.-H. Wang, K.~Han,
  J.~A. Jaramillo-Villegas, A.~{Al Noman}, C.~Wang, S.~Kim, M.~Teng, Y.~J. Lee,
  B.~Niu, L.~Fan, J.~Wang, D.~E. Leaird, A.~M. Weiner, and M.~Qi,
  \enquote{{High-Q silicon nitride microresonators exhibiting low-power
  frequency comb initiation},} {\protect\JournalTitle{Optica}} \textbf{3}, 1171
  (2016).

\bibitem{Lu2019a}
X.~Lu, Q.~Li, D.~A. Westly, G.~Moille, A.~Singh, V.~Anant, and K.~Srinivasan,
  \enquote{{Chip-integrated visible–telecom entangled photon pair source for
  quantum communication},} {\protect\JournalTitle{Nature Physics}} \textbf{15},
  373--381 (2019).

\bibitem{Elshaari2020}
A.~W. Elshaari, W.~Pernice, K.~Srinivasan, O.~Benson, and V.~Zwiller,
  \enquote{{Hybrid integrated quantum photonic circuits},}
  {\protect\JournalTitle{Nature Photonics}} \textbf{14}, 285--298 (2020).

\bibitem{Kim2020}
J.~J.-H. Kim, S.~Aghaeimeibodi, J.~Carolan, D.~Englund, and E.~Waks,
  \enquote{{Hybrid integration methods for on-chip quantum photonics},}
  {\protect\JournalTitle{Optica}} \textbf{7}, 291 (2020).

\bibitem{Lake2016}
D.~P. Lake, M.~Mitchell, H.~Jayakumar, L.~F. dos Santos, D.~Curic, and P.~E.
  Barclay, \enquote{{Efficient telecom to visible wavelength conversion in
  doubly resonant gallium phosphide microdisks},}
  {\protect\JournalTitle{Applied Physics Letters}} \textbf{108}, 031109 (2016).

\bibitem{Wilson2020}
D.~J. Wilson, K.~Schneider, S.~H{\"{o}}nl, M.~Anderson, Y.~Baumgartner,
  L.~Czornomaz, T.~J. Kippenberg, and P.~Seidler, \enquote{{Integrated gallium
  phosphide nonlinear photonics},} {\protect\JournalTitle{Nature Photonics}}
  \textbf{14}, 57--62 (2020).

\bibitem{Jung2021}
H.~Jung, S.-P. Yu, D.~R. Carlson, T.~E. Drake, T.~C. Briles, and S.~B. Papp,
  \enquote{{Tantala Kerr nonlinear integrated photonics},}
  {\protect\JournalTitle{Optica}} \textbf{8}, 811 (2021).

\bibitem{Guidry2020}
M.~A. Guidry, K.~Y. Yang, D.~M. Lukin, A.~Markosyan, J.~Yang, M.~M. Fejer, and
  J.~Vu{\v{c}}kovi{\'{c}}, \enquote{{Optical parametric oscillation in silicon
  carbide nanophotonics},} {\protect\JournalTitle{Optica}} \textbf{7}, 1139
  (2020).

\bibitem{Wang2021}
C.~Wang, Z.~Fang, A.~Yi, B.~Yang, Z.~Wang, L.~Zhou, C.~Shen, Y.~Zhu, Y.~Zhou,
  R.~Bao, Z.~Li, Y.~Chen, K.~Huang, J.~Zhang, Y.~Cheng, and X.~Ou,
  \enquote{{High-Q microresonators on 4H-silicon-carbide-on-insulator platform
  for nonlinear photonics},} {\protect\JournalTitle{Light: Science \&
  Applications}} \textbf{10}, 139 (2021).

\bibitem{Wu2018}
R.~Wu, J.~Zhang, N.~Yao, W.~Fang, L.~Qiao, Z.~Chai, J.~Lin, and Y.~Cheng,
  \enquote{{Lithium niobate micro-disk resonators of quality factors above
  $10^7$},} {\protect\JournalTitle{Optics Letters}} \textbf{43}, 4116 (2018).

\bibitem{Zhu2021}
D.~Zhu, L.~Shao, M.~Yu, R.~Cheng, B.~Desiatov, C.~J. Xin, Y.~Hu, J.~Holzgrafe,
  S.~Ghosh, A.~Shams-Ansari, E.~Puma, N.~Sinclair, C.~Reimer, M.~Zhang, and
  M.~Lon{\v{c}}ar, \enquote{{Integrated photonics on thin-film lithium
  niobate},} {\protect\JournalTitle{Advances in Optics and Photonics}}
  \textbf{13}, 242 (2021).

\bibitem{wang2018nanophotonic}
C.~Wang, M.~Zhang, B.~Stern, M.~Lipson, and M.~Lon{\v{c}}ar,
  \enquote{Nanophotonic lithium niobate electro-optic modulators,}
  {\protect\JournalTitle{Optics express}} \textbf{26}, 1547--1555 (2018).

\bibitem{zhang2021integrated}
M.~Zhang, C.~Wang, P.~Kharel, D.~Zhu, and M.~Lon{\v{c}}ar, \enquote{Integrated
  lithium niobate electro-optic modulators: when performance meets
  scalability,} {\protect\JournalTitle{Optica}} \textbf{8}, 652--667 (2021).

\bibitem{mckenna2020cryogenic}
T.~P. McKenna, J.~D. Witmer, R.~N. Patel, W.~Jiang, R.~Van~Laer,
  P.~Arrangoiz-Arriola, E.~A. Wollack, J.~F. Herrmann, and A.~H. Safavi-Naeini,
  \enquote{Cryogenic microwave-to-optical conversion using a triply resonant
  lithium-niobate-on-sapphire transducer,} {\protect\JournalTitle{Optica}}
  \textbf{7}, 1737--1745 (2020).

\bibitem{holzgrafe2020cavity}
J.~Holzgrafe, N.~Sinclair, D.~Zhu, A.~Shams-Ansari, M.~Colangelo, Y.~Hu,
  M.~Zhang, K.~K. Berggren, and M.~Lon{\v{c}}ar, \enquote{Cavity electro-optics
  in thin-film lithium niobate for efficient microwave-to-optical
  transduction,} {\protect\JournalTitle{Optica}} \textbf{7}, 1714--1720 (2020).

\bibitem{jiang2019lithium}
W.~Jiang, R.~N. Patel, F.~M. Mayor, T.~P. McKenna, P.~Arrangoiz-Arriola, C.~J.
  Sarabalis, J.~D. Witmer, R.~Van~Laer, and A.~H. Safavi-Naeini,
  \enquote{Lithium niobate piezo-optomechanical crystals,}
  {\protect\JournalTitle{Optica}} \textbf{6}, 845--853 (2019).

\bibitem{jiang2022optically}
W.~Jiang, F.~M. Mayor, S.~Malik, R.~Van~Laer, T.~P. McKenna, R.~N. Patel, J.~D.
  Witmer, and A.~H. Safavi-Naeini, \enquote{Optically heralded microwave
  photons,} {\protect\JournalTitle{arXiv preprint arXiv:2210.10739}}  (2022).

\bibitem{mckenna2022ultra}
T.~P. McKenna, H.~S. Stokowski, V.~Ansari, J.~Mishra, M.~Jankowski, C.~J.
  Sarabalis, J.~F. Herrmann, C.~Langrock, M.~M. Fejer, and A.~H. Safavi-Naeini,
  \enquote{Ultra-low-power second-order nonlinear optics on a chip,}
  {\protect\JournalTitle{Nature Communications}} \textbf{13}, 4532 (2022).

\bibitem{park2022high}
T.~Park, H.~S. Stokowski, V.~Ansari, T.~P. McKenna, A.~Y. Hwang, M.~Fejer, and
  A.~H. Safavi-Naeini, \enquote{High-efficiency second harmonic generation of
  blue light on thin-film lithium niobate,} {\protect\JournalTitle{Optics
  Letters}} \textbf{47}, 2706--2709 (2022).

\bibitem{Mouradian2017}
S.~Mouradian, N.~H. Wan, T.~Schr{\"{o}}der, and D.~Englund,
  \enquote{{Rectangular photonic crystal nanobeam cavities in bulk diamond},}
  {\protect\JournalTitle{Applied Physics Letters}} \textbf{111}, 021103 (2017).

\bibitem{Zhu2022}
Y.~Zhu, W.~Wei, A.~Yi, T.~Jin, C.~Shen, X.~Wang, L.~Zhou, C.~Wang, W.~Ou,
  S.~Song, T.~Wang, J.~Zhang, X.~Ou, and J.~Zhang, \enquote{{Hybrid Integration
  of Deterministic Quantum Dot‐Based Single‐Photon Sources with
  CMOS‐Compatible Silicon Carbide Photonics},} {\protect\JournalTitle{Laser
  \& Photonics Reviews}} \textbf{16}, 2200172 (2022).

\bibitem{Chanana2022}
A.~Chanana, H.~Larocque, R.~Moreira, J.~Carolan, B.~Guha, V.~Anant, J.~D. Song,
  D.~Englund, D.~J. Blumenthal, K.~Srinivasan, and M.~Davanco,
  \enquote{{Triggered single-photon generation and resonance fluorescence in
  ultra-low loss integrated photonic circuits},} {\protect\JournalTitle{arXiv:
  2202.04615}}  (2022).

\bibitem{Xie2018}
L.~Xie, T.~X. Zhou, R.~J. St{\"{o}}hr, and A.~Yacoby,
  \enquote{{Crystallographic Orientation Dependent Reactive Ion Etching in
  Single Crystal Diamond},} {\protect\JournalTitle{Advanced Materials}}
  \textbf{30}, 1705501 (2018).

\bibitem{Khanaliloo2015}
B.~Khanaliloo, M.~Mitchell, A.~C. Hryciw, and P.~E. Barclay, \enquote{{High- Q
  / V Monolithic Diamond Microdisks Fabricated with Quasi-isotropic Etching},}
  {\protect\JournalTitle{Nano Letters}} \textbf{15}, 5131--5136 (2015).

\bibitem{Khanaliloo2015b}
B.~Khanaliloo, H.~Jayakumar, A.~C. Hryciw, D.~P. Lake, H.~Kaviani, and P.~E.
  Barclay, \enquote{{Single-Crystal Diamond Nanobeam Waveguide Optomechanics},}
  {\protect\JournalTitle{Physical Review X}} \textbf{5}, 041051 (2015).

\bibitem{Mouradian2015}
S.~L. Mouradian, T.~Schr{\"{o}}der, C.~B. Poitras, L.~Li, J.~Goldstein, E.~H.
  Chen, M.~Walsh, J.~Cardenas, M.~L. Markham, D.~J. Twitchen, M.~Lipson, and
  D.~Englund, \enquote{{Scalable Integration of Long-Lived Quantum Memories
  into a Photonic Circuit},} {\protect\JournalTitle{Physical Review X}}
  \textbf{5}, 031009 (2015).

\bibitem{Nguyen2019a}
C.~T. Nguyen, D.~D. Sukachev, M.~K. Bhaskar, B.~Machielse, D.~S. Levonian,
  E.~N. Knall, P.~Stroganov, C.~Chia, M.~J. Burek, R.~Riedinger, H.~Park,
  M.~Lon{\v{c}}ar, and M.~D. Lukin, \enquote{{An integrated nanophotonic
  quantum register based on silicon-vacancy spins in diamond},}
  {\protect\JournalTitle{Physical Review B}} \textbf{100}, 165428 (2019).

\bibitem{Rugar2020}
A.~E. Rugar, H.~Lu, C.~Dory, S.~Sun, P.~J. McQuade, Z.-X. Shen, N.~A. Melosh,
  and J.~Vu{\v{c}}kovi{\'{c}}, \enquote{{Generation of Tin-Vacancy Centers in
  Diamond via Shallow Ion Implantation and Subsequent Diamond Overgrowth},}
  {\protect\JournalTitle{Nano Letters}} \textbf{20}, 1614--1619 (2020).

\bibitem{Schroder2017}
T.~Schr{\"{o}}der, M.~E. Trusheim, M.~Walsh, L.~Li, J.~Zheng, M.~Schukraft,
  A.~Sipahigil, R.~E. Evans, D.~D. Sukachev, C.~T. Nguyen, J.~L. Pacheco, R.~M.
  Camacho, E.~S. Bielejec, M.~D. Lukin, and D.~Englund, \enquote{{Scalable
  focused ion beam creation of nearly lifetime-limited single quantum emitters
  in diamond nanostructures},} {\protect\JournalTitle{Nature Communications}}
  \textbf{8}, 15376 (2017).

\bibitem{Titze2022}
M.~Titze, H.~Byeon, A.~Flores, J.~Henshaw, C.~T. Harris, A.~M. Mounce, and
  E.~S. Bielejec, \enquote{{In Situ Ion Counting for Improved Implanted Ion
  Error Rate and Silicon Vacancy Yield Uncertainty},}
  {\protect\JournalTitle{Nano Letters}} \textbf{22}, 3212--3218 (2022).

\bibitem{Rugar2021a}
A.~E. Rugar, S.~Aghaeimeibodi, D.~Riedel, C.~Dory, H.~Lu, P.~J. McQuade, Z.-X.
  Shen, N.~A. Melosh, and J.~Vu{\v{c}}kovi{\'{c}}, \enquote{{Quantum Photonic
  Interface for Tin-Vacancy Centers in Diamond},}
  {\protect\JournalTitle{Physical Review X}} \textbf{11}, 031021 (2021).

\bibitem{Nguyen2019b}
C.~T. Nguyen, D.~D. Sukachev, M.~K. Bhaskar, B.~Machielse, D.~S. Levonian,
  E.~N. Knall, P.~Stroganov, R.~Riedinger, H.~Park, M.~Lon{\v{c}}ar, and M.~D.
  Lukin, \enquote{{Quantum Network Nodes Based on Diamond Qubits with an
  Efficient Nanophotonic Interface},} {\protect\JournalTitle{Physical Review
  Letters}} \textbf{123}, 183602 (2019).

\bibitem{Sutula2022}
M.~Sutula, I.~Christen, E.~Bersin, M.~P. Walsh, K.~C. Chen, J.~Mallek,
  A.~Melville, M.~Titze, E.~S. Bielejec, S.~Hamilton, D.~Braje, P.~B. Dixon,
  and D.~R. Englund, \enquote{{Large-scale optical characterization of
  solid-state quantum emitters},} {\protect\JournalTitle{arXiv: 2210.13643}}
  (2022).

\bibitem{Narita2022}
Y.~Narita, P.~Wang, K.~Oba, Y.~Miyamoto, T.~Taniguchi, S.~Onoda, M.~Hatano, and
  T.~Iwasaki, \enquote{{Identical Photons from Multiple Tin-Vacancy Centers in
  Diamond},} {\protect\JournalTitle{arXiv: 2208.0627}}  (2022).

\bibitem{Elshaari2017}
A.~W. Elshaari, I.~E. Zadeh, A.~Fognini, M.~E. Reimer, D.~Dalacu, P.~J. Poole,
  V.~Zwiller, and K.~D. J{\"{o}}ns, \enquote{{On-chip single photon filtering
  and multiplexing in hybrid quantum photonic circuits},}
  {\protect\JournalTitle{Nature Communications}} \textbf{8}, 379 (2017).

\bibitem{Reithmaier2015}
G.~Reithmaier, M.~Kaniber, F.~Flassig, S.~Lichtmannecker, K.~M{\"{u}}ller,
  A.~Andrejew, J.~Vu{\v{c}}kovi{\'{c}}, R.~Gross, and J.~J. Finley,
  \enquote{{On-Chip Generation, Routing, and Detection of Resonance
  Fluorescence},} {\protect\JournalTitle{Nano Letters}} \textbf{15}, 5208--5213
  (2015).

\bibitem{Najafi2015}
F.~Najafi, J.~Mower, N.~C. Harris, F.~Bellei, A.~Dane, C.~Lee, X.~Hu,
  P.~Kharel, F.~Marsili, S.~Assefa, K.~K. Berggren, and D.~Englund,
  \enquote{{On-chip detection of non-classical light by scalable integration of
  single-photon detectors},} {\protect\JournalTitle{Nature Communications}}
  \textbf{6}, 5873 (2015).

\bibitem{zeng2023fiberintegration}
B.~Zeng, C.~De-Eknamkul, D.~Assumpcao, D.~Renaud, Z.~Wang, D.~Riedel, J.~Ha,
  C.~Robens, D.~Levonian, M.~Lukin, M.~Bhaskar, D.~Sukachev, M.~Loncar, and
  B.~Machielse, \enquote{Optically heralded microwave photons,}
  {\protect\JournalTitle{arXiv preprint arXiv:2306.09894}}  (2023).

\bibitem{larocque2023tunable}
H.~Larocque, M.~A. Buyukkaya, C.~Errando-Herranz, S.~Harper, J.~Carolan, C.-M.
  Lee, C.~J.~K. Richardson, G.~L. Leake, D.~J. Coleman, M.~L. Fanto, E.~Waks,
  and D.~Englund, \enquote{Tunable quantum emitters on large-scale foundry
  silicon photonics,}  (2023).

\end{thebibliography}

\newpage
\title{Supplementary Material}
\setcounter{section}{0}
\section{Additional Methods and Calibrations}\label{secA}
The sample was maintained at 5 K in a Montana Instruments Cryostation s50. We position the sample using a three-axis stage consisting of an Attocube piezo stack (X101/Z100). We use a heated cryo-objective inside the Cryostation to address the sample. For grating coupler transmission measurements in Fig. 3, light couples in and out of the path via single-mode fibers. These are coupled into the path using fiber couplers with f=8.00 mm aspheric lenses (Thorlabs, C240TMD-B). For confocal excitation and collection, we again couple light into the path via fiber couplers, installed with f=18.4 mm aspheric lenses (Thorlabs, C280TMD-B). Our input and output collection fibers for all paths are 630 nm polarization-maintaining single mode fibers (Thorlabs, P3-630PM-FC-2). The excitation laser light (MSquared, Solstis) is first routed through a 3-paddle polarization controller (Thorlabs, FPC560) fitted with a single mode 630 nm fiber (Thorlabs, P3-630Y-FC-2). From here it is passed through a 99:1 single mode fiber beamsplitter (Thorlabs, TW670R1A2), which is used to monitor input power. 

We filter extraneous fiber fluorescence on the excitation path with a 711/25 nm (Semrock Brightline 711/25) bandpass filter and on all collection paths with 740/13 nm (Semrock Brightline 740/13) bandpass filters. 

Our confocal excitation/collection path consists of a 4f lens setup containing a fast steering mirror (Newport, FSM-300-01) and two f=300 mm achromatic doublets (Thorlabs, AC254-300-A). The confocal and coupler transmission paths are mixed via 720 nm shortpas dichroic beamsplitters (Semrock, FF720-SDi01-25x36). To compensate for polarization misalignment between the two dichroics, we use a 400-800 nm half-wave plate (Thorlabs, AHWP10M-600) to optimize power delivery. Final polarization control before the Montana window is achieved by a 690-1200 nm half-wave plate (Thorlabs, AQWP05M-980). All mirrors in the optical setup are broadband silver mirrors (Thorlabs, PF10-03-P01). Color center emitter signal is either routed to avalanche photodiode single photon counting modules (Excelitas, SPCM-AQRH-24) or a hybrid spectrometer with Acton, SpectraPro 2750 gratings and an Andor, iDus416 CCD. Power measurements are performed with photodiode sensors (Thorlabs, S120C) moved between parts of the optical path.

For g$^{(2)}$ autocorrelation measurements collected via the confocal channel, the signal is routed to a 50:50 fiber beamsplitter. Each arm is subsequently routed to a separate SPCM unit. For coupler-collected autocorrelation measurements, the output from each LN grating coupler is directly routed to separate SPCM, utilizing the TFLN waveguide intrinsically as a beamsplitter. SPCM outputs are then routed to a Picoharp 300 timetagging system. For spectrometer measurements, we route the optical signal through a 750/10 nm bandpass (Thorlabs FBH750-10) and a 750 nm shortpass (Thorlabs FES0750) in series, thereby filtering out the specific SiV- optical transitions we would like to measure. We tilt both filters to tune the filtering range and effectively isolate a single transition for PL detection. 

From manufacturer-provided specifications of the cryostat, cryo-objective, and mirrors, we assume a 95\% transmission of the cryostat window, 90\% through the cryo-objective, and 98\% reflection for each silver mirror. Lenses are assumed to have a transmission of 99.75\% each. SPCMs are assumed to have 65\% efficiency at operating wavelengths. 

\section{Diamond Nanobeam Fabrication}\label{secB}
We begin sample preparation with a 2 by 2 mm$^2$ electronic grade diamond from Element 6. The sample is cleaned with a refluxing triacid mixture, followed by soaking in acetone and isopropanol (IPA). We remove 500 nm of the strained diamond surface with an anisotropic, oxygen plasma etch. We incorporate SiV centers by growing back 200 nm of diamond, while placing a silicon substrate alongside the diamond in the CVD chamber. 

A hardmask consisting of 200 nm of Si$_{x}$N$_{y}$ is then deposited via low pressure CVD. The nanobeam pattern is defined in ZEP 520A positive ebeam resist with a 100keV ebeam writer (JEOL 6300) and developed in a three step process (ortho-xylene, MIBK:IPA 1:3, and IPA). We use an anisotropic etch with SF$_{6}$, CH$_{4}$, and N$_{2}$ to transfer the pattern from the resist to the hardmask, which is subsequently transferred into diamond with a 500 nm anisotropic oxygen plasma etch. Approximately 25 nm of Al$_2$O$_3$ is then deposited with atomic layer deposition (ALD) to protect the device sidewalls. A breakthrough etch of the alumina is performed with an ICP metal etcher tool (Plasma Therm Versaline LL ICP Metal Etcher). An additional 200 nm anisotropic diamond etch then exposes material, enabling continuation of a quasi-isotropic etch. 

The quasi-isotropic etch process is performed at an elevated temperature of 300 C using high density oxygen plasma in an ICP dielectric etch tool (Plasma Therm Versaline LL ICP Dielectric Etcher). We periodically remove the sample from the etcher and image with an SEM in order to monitor the progress of the undercut. Upon completion, we strip the masking layers with an HF soak. We then rinse the sample in water followed by IPA and allow it to dry by letting the solvent evaporate gradually.

\section{Pick and Place Details}\label{secC}
\begin{figure}[t!]
\centering
\includegraphics[width=\textwidth]{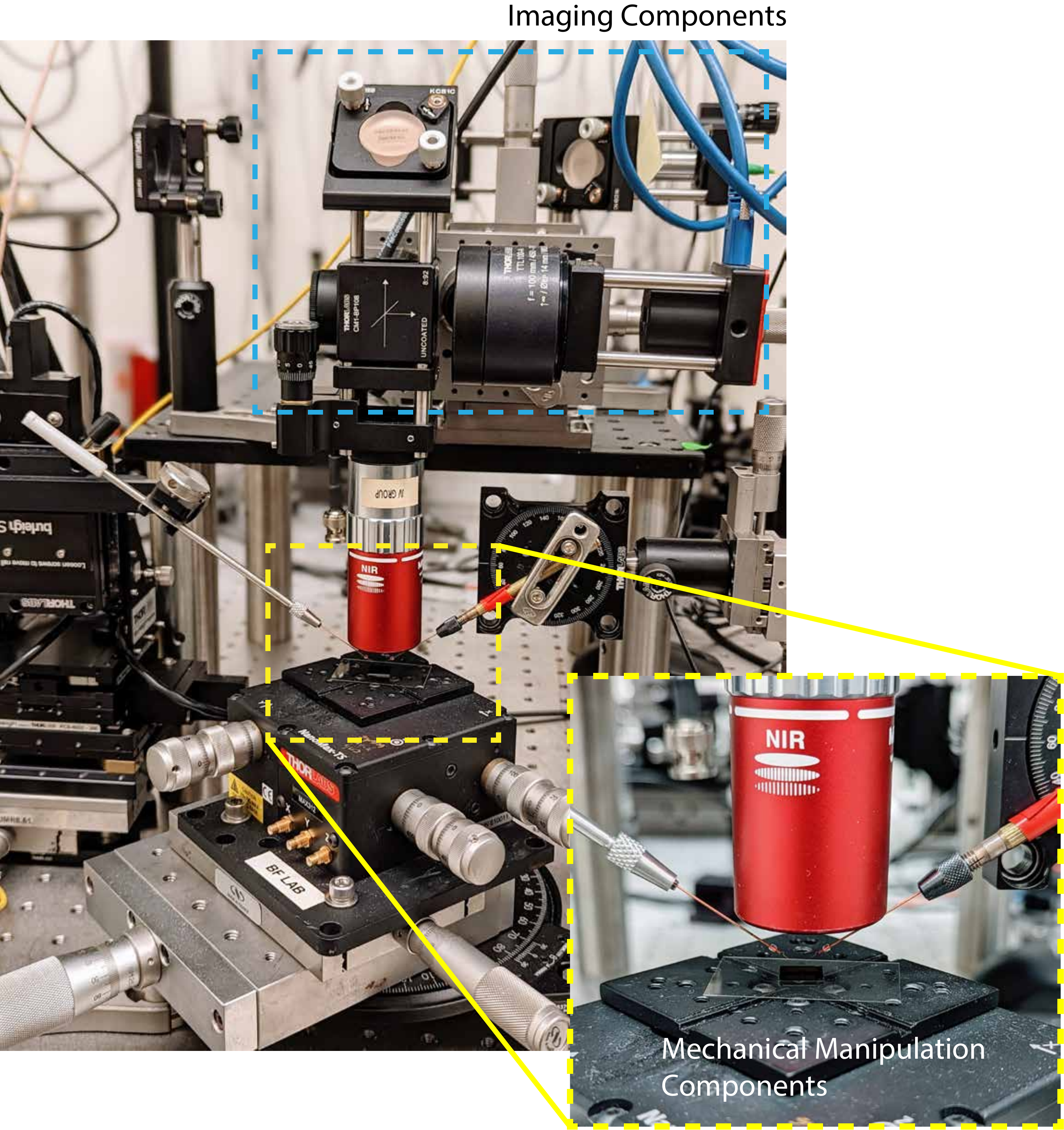}
\caption{Photo of pick and place setup. The mechanical manipulation components are highlighted by the yellow box, with the inset showing the components in more detail. The blue box indicates the imaging optic components.
} 
\label{fig:supfig0}
\end{figure}

In Fig.~\ref{fig:supfig0}, we present a photo of the pick and place setup. In order to monitor the transfer process \textit{in situ}, we use a long working distance object with WD$\sim 30$mm (Mitutoyo 100x M Plan APO NIR Objective - 378-826-15). 

One needle is a piezo-controlled (Thorlabs PCS-6300CL) tungsten ``cat whisker” needles with $\sim 70$\,nm tip radius (Ted Pella 99-15864) for initial break-off picking. To assist with the mechanical break-off process, some waveguides are detached from their diamond holding structures on the diamond substrate through a fixed ion beam (FIB) cut. 

In addition to a fine-controlled needle, a secondary cat whisker needle is mounted on a micrometer-controlled stage to provide an extra point of contact that enables mid-transfer control and adjustment of the diamond waveguide. Orientation prior to attachment of the diamond waveguide is optimized with the two needles and a rotation stage. The waveguide is transferred to the fine-controlled needle and lowered to the LN substrate. After making contact with the LN gratings, the diamond attaches electrostatically more strongly to the LN than to the needle, and the needle is further pulled down into an etched trench between the LN gratings as it detaches from the diamond. Further fine position and alignment adjustment can be done with the needle while taking care not to apply force sufficient to break the diamond from LN.

\section{Efficiency Calculations}\label{secD}
We indicate the taper facet transfer efficiency to be $\eta_{\rm taper}$, the grating coupler efficiency to be $\eta_{\rm coupler}$, and all other optical losses lumped together to be $\eta_{\rm losses}$. Therefore, our measured transmissions (T) can be expressed as:
\begin{equation}
    T = \eta_{\rm taper1}\eta_{\rm taper2}\eta_{\rm coupler1} \eta_{\rm coupler2}\eta_{\rm losses}
\end{equation}
We then make the reasonable assumption:
\begin{equation}
    \eta_{\rm taper1}\simeq\eta_{\rm taper2}=\eta_{\rm taper}
\end{equation}
We lump the grating coupler input and output efficiencies into a single term (see supplemental section \ref{fig:supplement_g} for a discussion of the expected in- and out-coupling efficiencies):
\begin{equation}
    \eta_{\rm coupler1}\eta_{\rm coupler2}=\eta_{\rm coupler}
\end{equation}
\newline
Which simplifies our transmission to:
\begin{equation}
    T \simeq \eta_{\rm taper}^2\eta_{\rm coupler}\eta_{\rm losses}
\end{equation}
We now make the approximation that the grating coupler efficiencies are roughly constant across the chip and therefore, $\eta_{\rm coupler, integrated}\simeq\eta_{\rm coupler, control}$. Additionally, as the two measurements were using identical optical paths, both are affected by the same $\eta_{\rm losses}$. Given that the control device consists of a continuous TFLN waveguide, $\eta_{\rm taper, control}=1$. Therefore, by comparing transmissions:
\begin{equation}
    T_{\rm integrated}/T_{\rm control}\simeq\eta_{\rm taper, integrated}^2
\end{equation}
\begin{equation}
    \eta_{\rm taper, integrated}\simeq\sqrt{T_{\rm integrated}/T_{\rm control}}
\end{equation}

\section{Measurement Calibrations}\label{secE}

\begin{figure}[t]
\includegraphics[width=\textwidth]{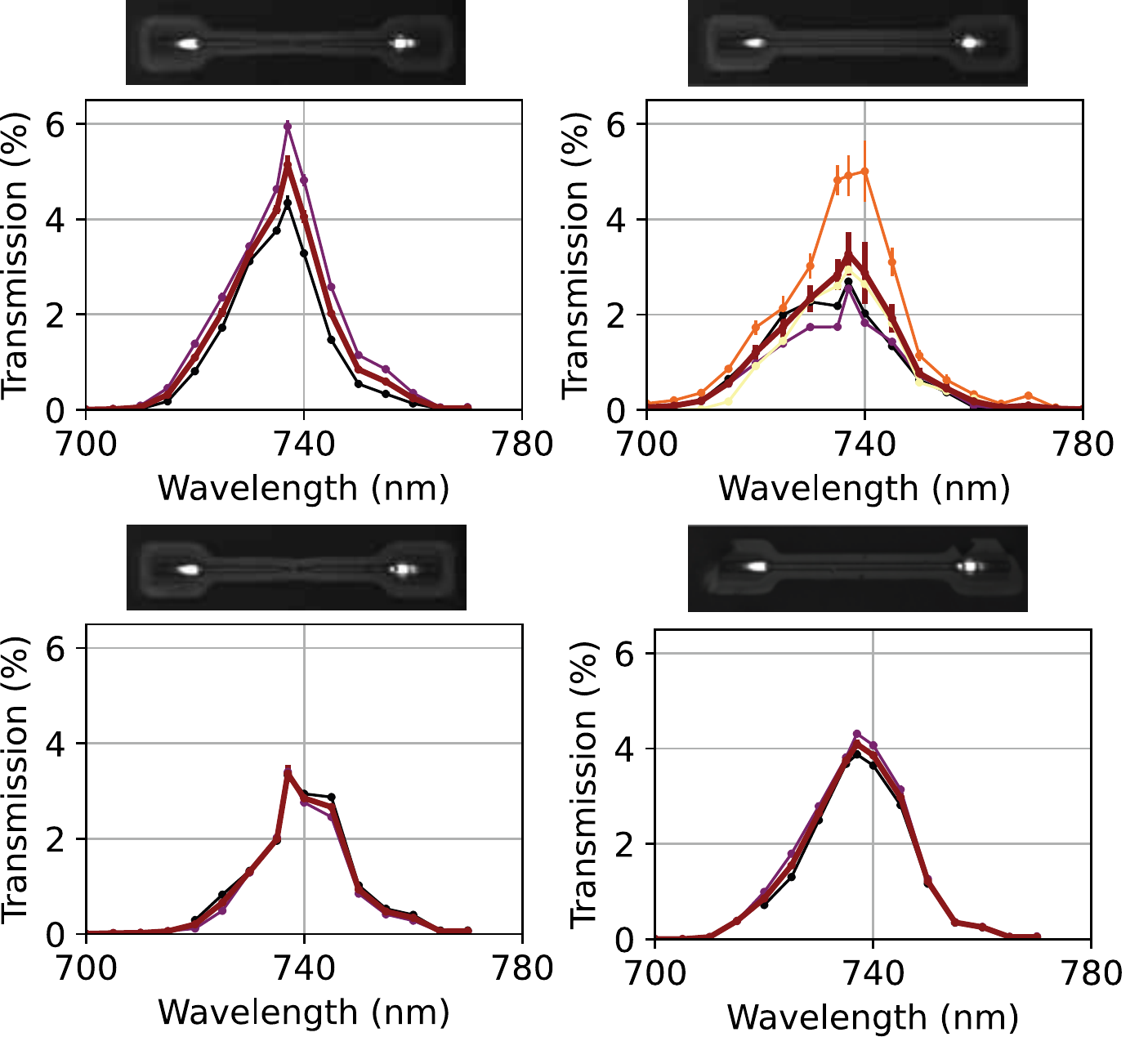}
\centering
\caption{Transmission measurements for control devices. Transmission measurements for the 4 control devices present on the TFLN chip, referred to as ``1'', ``2'', ``3'', and ``4'' in order from left to right, top to bottom. Different thin-line transmission traces are data from separate measurements, while the bold overlaid trace is the average across all measurements. We present data from multiple separate characterizations, taken over 2 months. Error bars indicate one standard deviation. For some data points, error bars are smaller than data markers. White light images of each device and its visible transmission are displayed above each transmission spectrum.
} 
\label{fig:supfig2a}
\end{figure}

\begin{figure}[t]
\includegraphics[width=\textwidth]{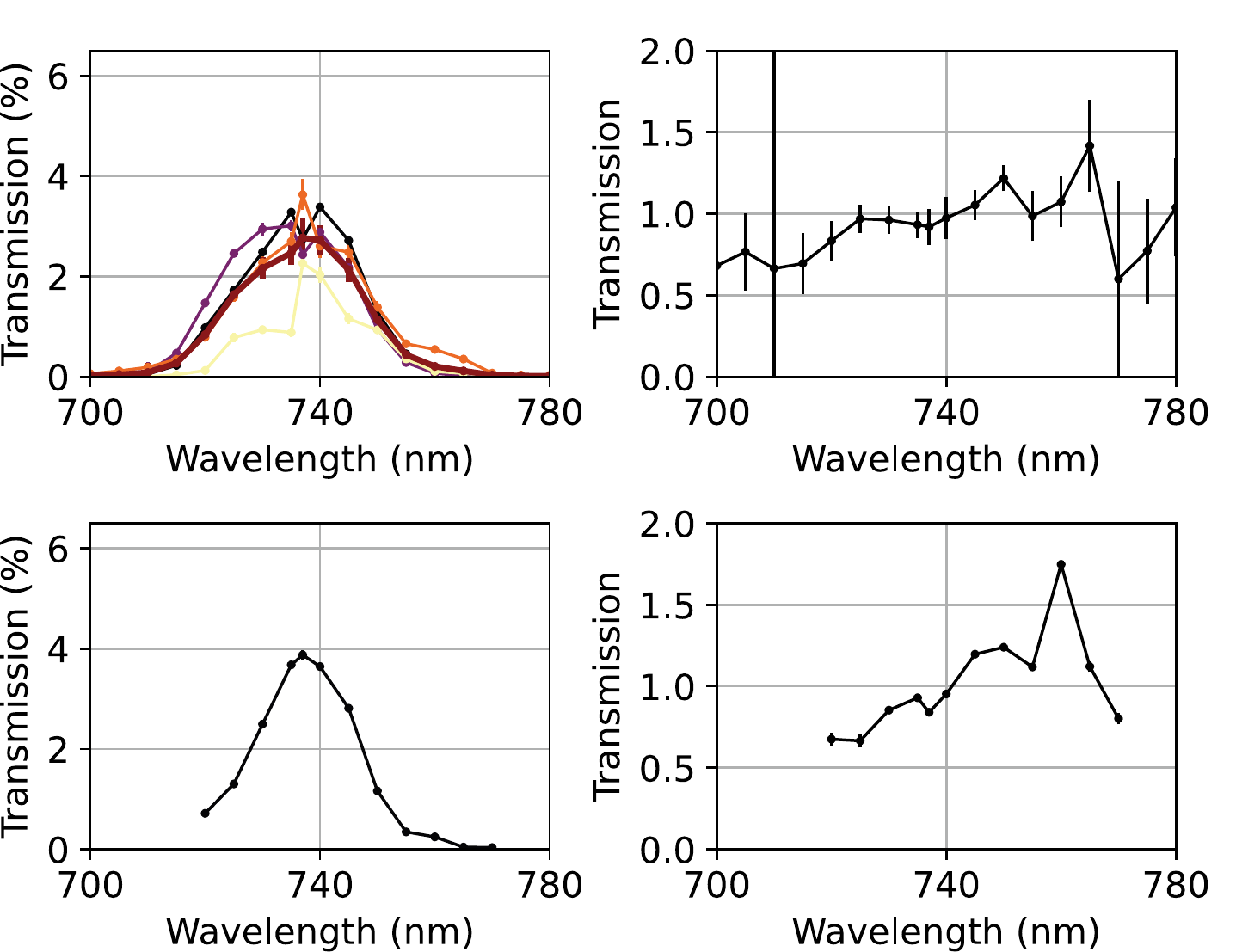}
\centering
\caption{Statistical evaluation of transmission measurements. Transmission data for the device reported in the manuscript (``Device 1'', top), and from an additional integrated device (``Device 2'', bottom) are presented. Transmission for device 1 was taken 4 separate times, over 2 months. For device 1, the thin transmission traces indicate data for each separate measurement, while the bold trace is the average across all measurements (left). The averaged ``Device 1'' transmission is normalized to the average transmission for control ``2'', generating the adiabatic taper efficiency response across wavelengths (right). Given very small transmission values at the extremities of the grating response for both integrated and control devices, taper efficiency error bars are observed to increase dramatically at these wavelengths. All error bars indicate one standard deviation. A similar measurement was taken for ``Device 2.''
} 
\label{fig:supfig2b}
\end{figure}

We fabricated four control devices for calibrating grating coupler efficiency on the TFLN sample. Two of these consist of TFLN tapered waveguides connected at the tips and two are constant-width waveguides. Transmission statistics were taken multiple times over 2 months for all four control devices as well as the heterogeneously integrated device. Transmission values are relatively consistent across multiple measurements, with any variance attributed to alignment or thermal fluctation, as well as gradual condensation in the cryosystem. Because the TFLN tapered waveguides introduce uncharacterized scattering, we only use the fully-connected constant-width waveguides for ``control device'' measurements. For measurements performed in the main text, we use connected device ``2'' as the control device. Connected device ``4'' was damaged during measurements and could not be used to calibrate transmission measurements. By measuring both devices, ``2'' and ``4'', we can estimate the total grating coupler in-to-out-coupling transmission at $737$ nm to range between $2.69\%$ and $4.92\%$, and averaging to $3.55\pm0.44\%$ (we assume negligible loss in the waveguide). This is pictured in Fig.~\ref{fig:supfig2a}. These values reasonably match the simulated in-to-out-coupling efficiencies of the grating couplers simulated via FDTD in section \ref{fig:supplement_g}.

From these transmission measurements, we determine the taper efficiency of a second fully-integrated device (different from that presented in the main text) to be $84.1\pm1.6\%$ per taper facet at $737$ nm, demonstrating the repeatability of our pick and place transfer procedure. These measurements are illustrated in Fig.~\ref{fig:supfig2b}.

\section{Emitter-Nanobeam Coupling Simulations}\label{secF}
To verify the validity of our confocal versus coupler collection channel count rates, we use Lumerical FDTD to perform simulated sweeps of dipole emitter orientation and position while monitoring transmission through the different collection channels. We simulate a nanobeam with $300$ nm width and $200$ nm thickness, modeling the single mode region of our tapered diamond devices. We place a monitor at the end of each nanobeam to observe transmission coupled into the TFLN waveguide, and a monitor directly above the emitter to collect out-of-plane, confocal scattering.

\begin{figure}[h!]
\centering
\includegraphics[width=\textwidth]{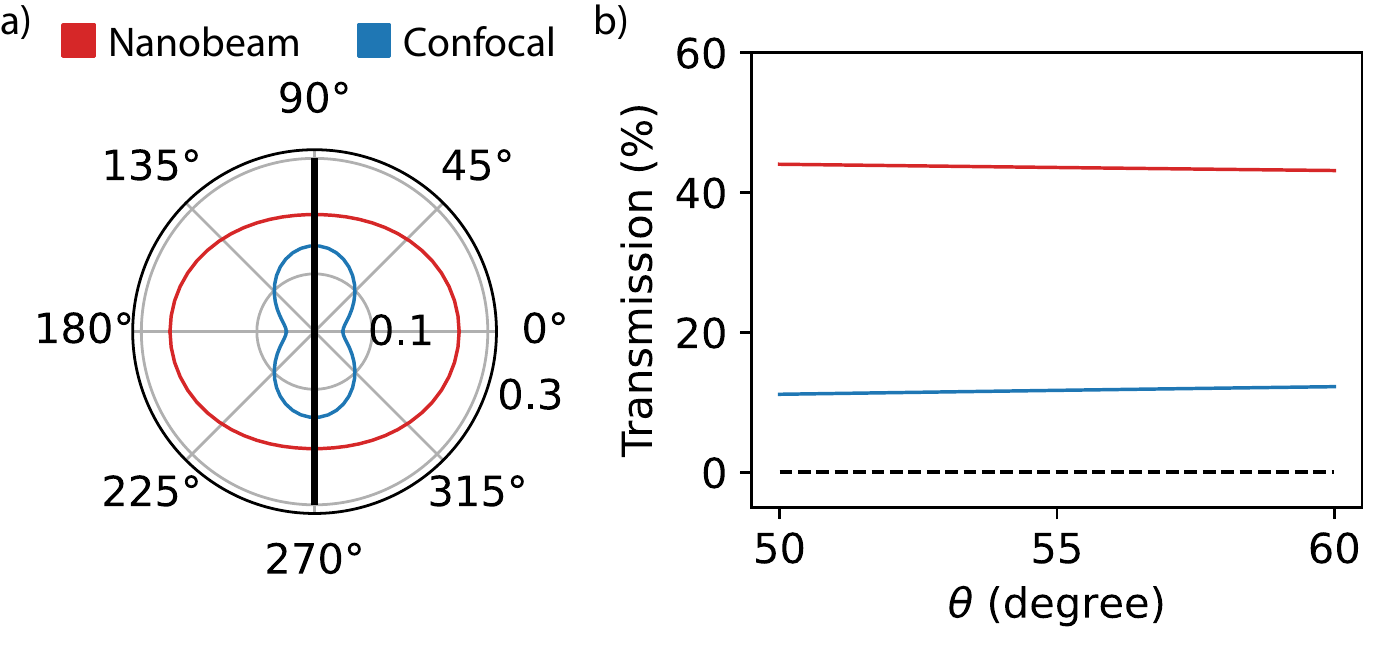}
\caption{Angular dipole orientation and nanobeam coupling.
\textbf{(a)} Transmission variance with fixed $\theta\approx55$\textdegree and sweeping full range of $ \phi$. Red corresponds to transmission collected through integrated channels, and blue confocal. In order to optimize photon extraction through integrated channels, we see that it is ideal to orient our dipole perpendicular to the nanobeam.
\textbf{(b)} Transmission variance with fixed $\phi\approx55$\textdegree and sweeping full range of $ \theta$. A small range of $ \theta$ is swept to take into account misalignment of cut of starting bulk crystals. We see that such small fluctuations in orientation do not cause great variation in either confocal or integration collection efficiencies, and that integrated collection channels consistently outperform confocal channels by nearly five-fold.
} 
\label{fig:supfig1}
\end{figure}

To investigate the effects of dipole orientation, we sweep the polar angle, $\theta$, of the dipole from $50$\textdegree to $60$\textdegree to account for slight crystalline misalignment out of plane. We sweep the azimuthal angle, $\phi$, between $0$\textdegree and $180$\textdegree, accounting for misalignment stemming from both the cut of the bulk crystal and lithography. For $\langle100\rangle$ cut diamond with devices perfectly aligned to the cubic axes, we expect both $\theta$ and $\phi$ to be $\sim 55$\textdegree, given diamond's tetrahedral structure.
Taking slices at both $\theta\approx55$\textdegree and $\phi\approx55$\textdegree and summing the transmitted power at both ends of the nanobeam, we see that in-nanobeam transmission (coupler channels) is greater than out-of-plane confocal collection, as shown in Fig.~\ref{fig:supfig1}.

\begin{figure}[t]
\centering
\includegraphics[width=\textwidth]{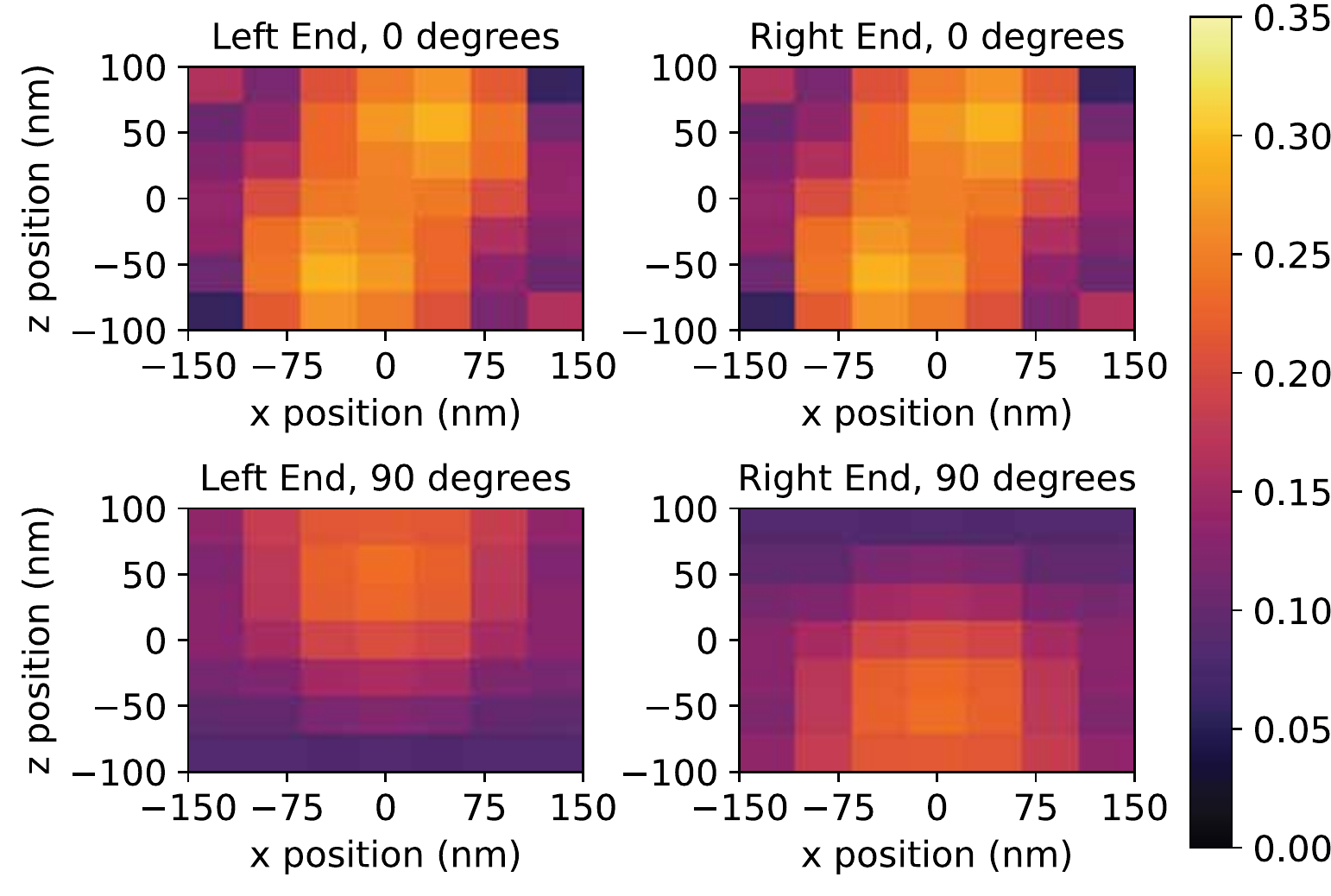}
\caption{Transmission efficiency at each nanobeam end as emitter position is varied in the device cross-section. Notably, depending on emitter positioning in the nanobeam, we observe an asymmetry in transmission efficiency at each device end.
} 
\label{fig:supfig1a}
\end{figure}

\begin{figure}[t]
\includegraphics[width=\textwidth]{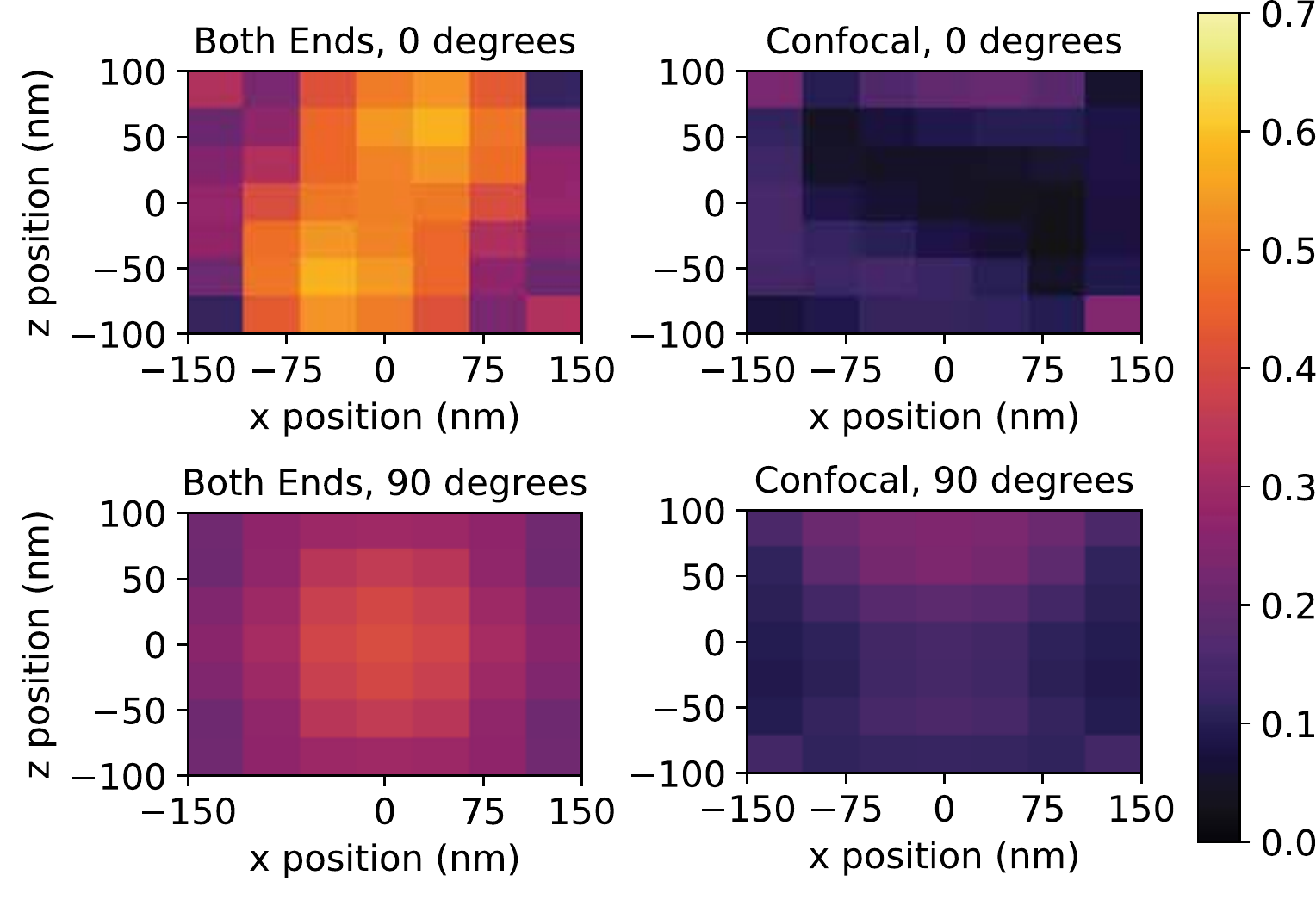}
\centering
\caption{Transmission efficiency as emitter position is varied in the nanobeam cross-section, comparing total signal collected through integrated channels to that confocally. We see that consistently across nearly all emitter positions, the integrated channels demonstrate improved photon collection rates.
} 
\label{fig:supfig1a}
\end{figure}

For sweeps of dipole position in the nanobeam cross-section, we maintain $ \phi=55$\textdegree. We vary $ \theta$ between $0$\textdegree and $90$\textdegree only, given these two angles correspond to maximum and minimum contrast between confocal and integrated transmission. Again, we observe that integrated collection channels consistently perform equally to or outperform the out-of-plane collection consistently across all emitter positions by nearly 5-fold Fig.\ref{fig:supfig1a}. We stress that because of realistic losses from optics and the angular dependence of the collection objective, we see instead an increased 10-fold improvement in experimental device demonstration.

\section{Grating Coupler Simulations}\label{secG}
\begin{figure}
   \centering
   \includegraphics[width=\textwidth]{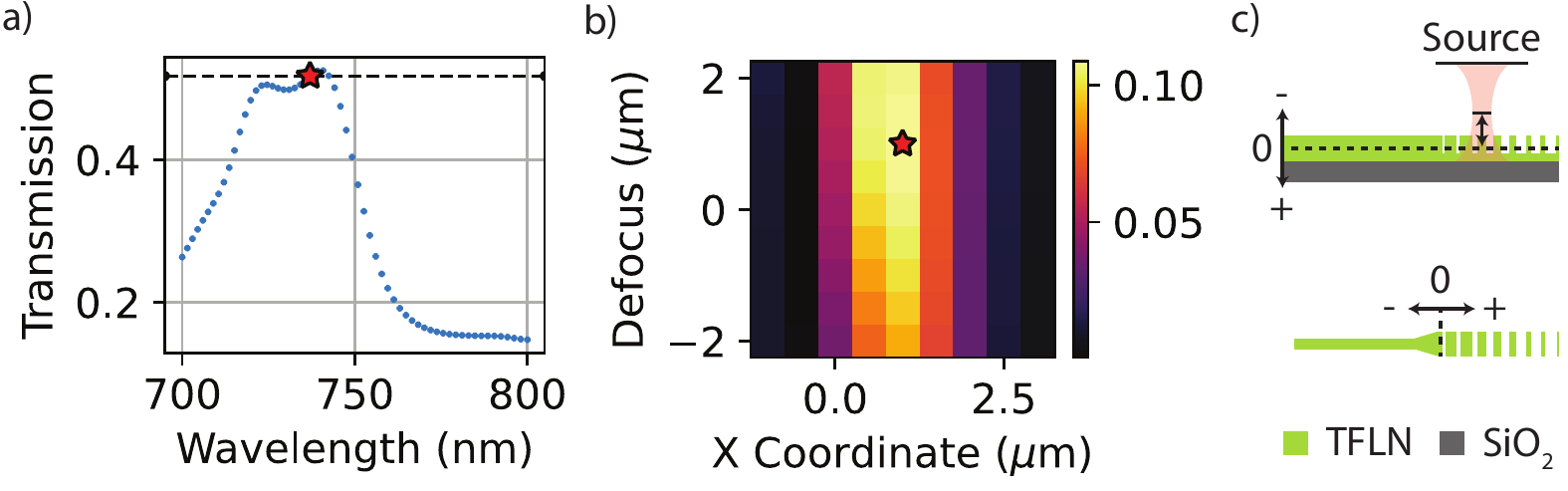}
   \caption{Simulated grating coupler response.
   \textbf{(a)} Simulated spectrum of the TFLN grating couplers used in the device. The simulation consists of a single-mode source in a waveguide and measures the transmitted power out of the grating into free space. The red star indicates an interpolated datapoint of transmission at 737 nm.
   \textbf{(b)} Simulated map of the single-mode in-coupling efficiency from a gaussian source into the TFLN waveguide. The ``x coordinate'' corresponds to position of the beam relative to the start of the grating, while the ``defocus'' parameter corresponds to vertical position relative to the center of the TFLN stack. The red star indicates the source position and defocus yielding maximum transmission into the waveguide fundamental mode.
   \textbf{(c)} Schematic of the device geometry used in simulation. (Top) Defocus parameter, indicating the distance of the source focus from the middle of the TFLN stack. (Bot) ``X Coordinate,'' indicating distance of the source focus from the start of the grating.}
\label{fig:supplement_g}
\end{figure}

In Fig.\ref{fig:supplement_g}, we demonstrate simulated design parameters for our grating couplers. Our couplers are approximately 2 $\mu$m wide and 10 $\mu$m long with a designed pitch in CAD of $425$ nm. The duty cycle is swept to improve coupling bandwidth. In practice, our actual devices and grating pitch differ a bit due to fabrication discrepancies from CAD. For these parameters, we simulate an expected out-coupling efficiency from a single-mode source in the waveguide of approximately 51.7\%, depicted in Fig.\ref{fig:supplement_g}(a). In-coupling of the light is more difficult, due to the multimoded behavior of the TFLN waveguide at 737 nm. Fig. \ref{fig:supplement_g}(b) displays a map of the simulated in-coupling efficiency into the waveguide fundamental mode as a function of the incident beam placement and focal plane. For the source placement corresponding to the maximum transmission from free space into the waveguide, we simulate an expected transmission into the fundamental waveguide mode of approximately $10.85\%$. Multiplying the single-mode input and output efficiencies together indicates that we can expect a single-mode in-to-out coupling through the simulated control device of approximately $5.6\%$, similar to our demonstrated experimental coupling. The full multimode transmission is expected to be slightly greater, and we expect these numeric results to upper-bound our experimentally measured grating efficiencies.  

\section{Power-Dependent Autocorrelation}\label{secH}
Each autocorrelation measurement presented in the main text is extracted from a series of power-dependent $g^{2}$ measurements. The full series of measurements are provided here, demonstrating how for both collection schemes, $g^{2}(0)$ does not fall below $0.5$, the single emitter threshold, but still demonstrates predominantly single emitter behavior of the interrogated spot. 

\begin{figure}[t]
\centering
\includegraphics[width=\textwidth]{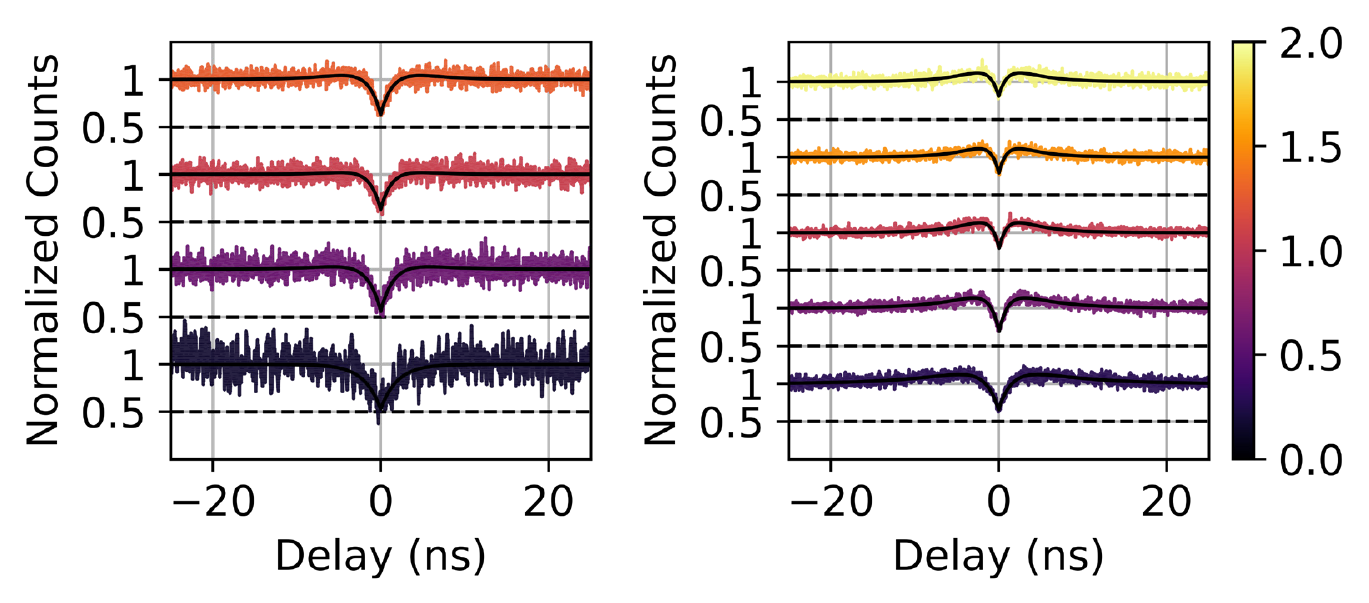}
\caption{Power-dependent autocorrelation series for confocal collection (left), and coupler collection (right). All $g^{(2)}(0)$ dips fail to fall below the 0.5 threshold, marked by a dashed line, which indicates that we are in fact probing a small ensemble of emitters. The power at which each curve is collected is expressed in factors of saturation power.
} 
\label{fig:supfig3a}
\end{figure}

\section{Focus-Dependent PL}\label{secI}

 We compare the integrated device's emission into a confocal collection channel to emission into the TFLN waveguide channel by varying the confocal excitation and collection foci and recording the resulting photoluminescence (PL) spectra measured via each collection channel. 
  
\begin{figure}[t]
\centering
\includegraphics[width=\textwidth]{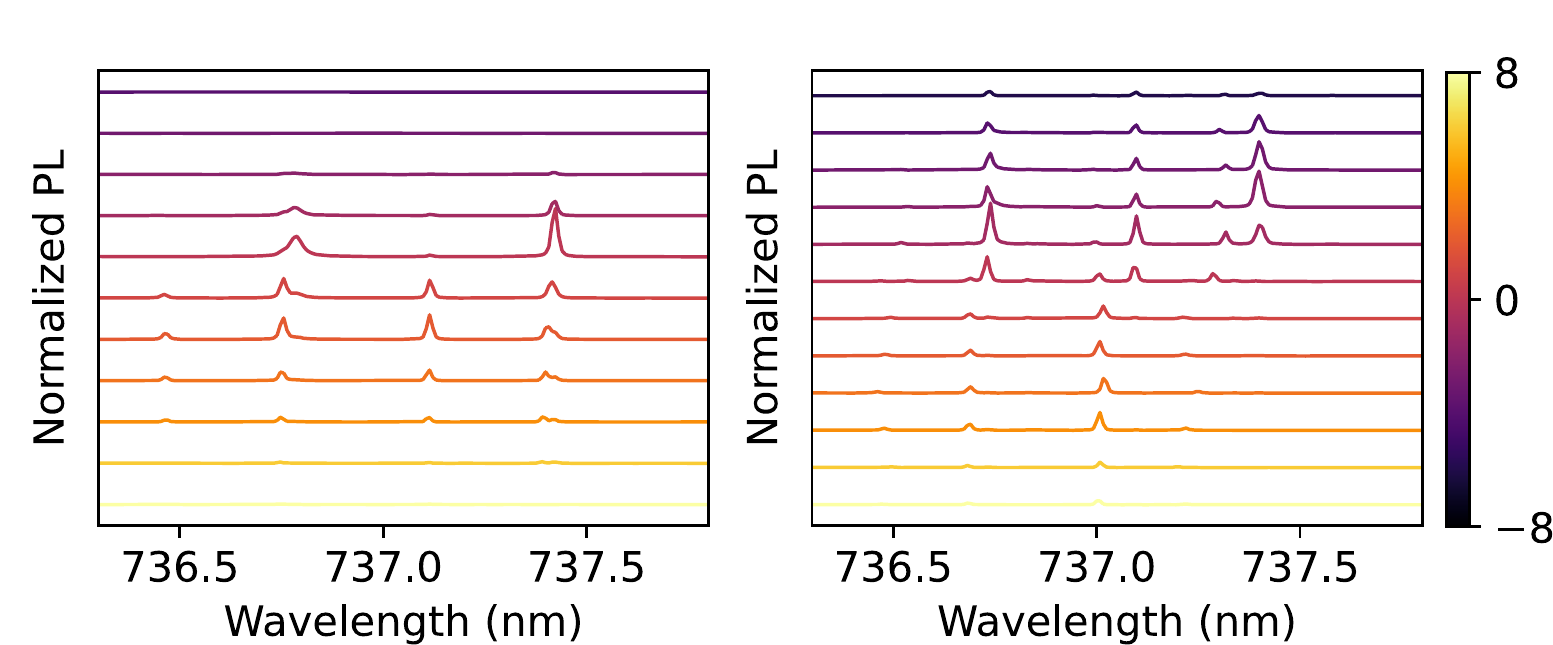}
\caption{Focus-dependent PL spectra. PL spectra obtained in the C/C configuration if displayed on left, and PL spectra obtained in the C/CP configuration, on right. Colorbar indicates number of rotations of the focal knob that was taken. As the excitation/collection plane is swept, the emitter transitions merge together on the spectrometer for the C/C configuration, appearing as a single emitter signature at certain focal planes. For C/P, we observe the spectral signatures of a greater number of emitters compared to the data for C/C.
} 
\label{fig:supfig3b}
\end{figure}

For the confocal excitation and collection scheme (C/C configuration), we step the foci in tandem, while localized to the same coordinate and plot the results in Fig.~\ref{fig:supfig3b}(a). While we observe emerging transitions at different foci, they mostly overlap in the C/C collection, maintaining the appearance of a ``single'' emitter at multiple focal planes. We predominantly observe a decrease in collected light as the device is moved out of focus.
 
 We repeat this measurement for the confocal excitation with the C/CP configuration. In this configuration, only the confocal excitation plane is stepped, while the collection plane is fixed on the TFLN grating couplers. We plot the results in Fig.~\ref{fig:supfig3b}(b).  Here, we more clearly observe multiple emitter transitions appearing at different foci, including a number not captured at all through confocal channels.
 
\begin{figure}[t]
\centering
\includegraphics[width=\textwidth]{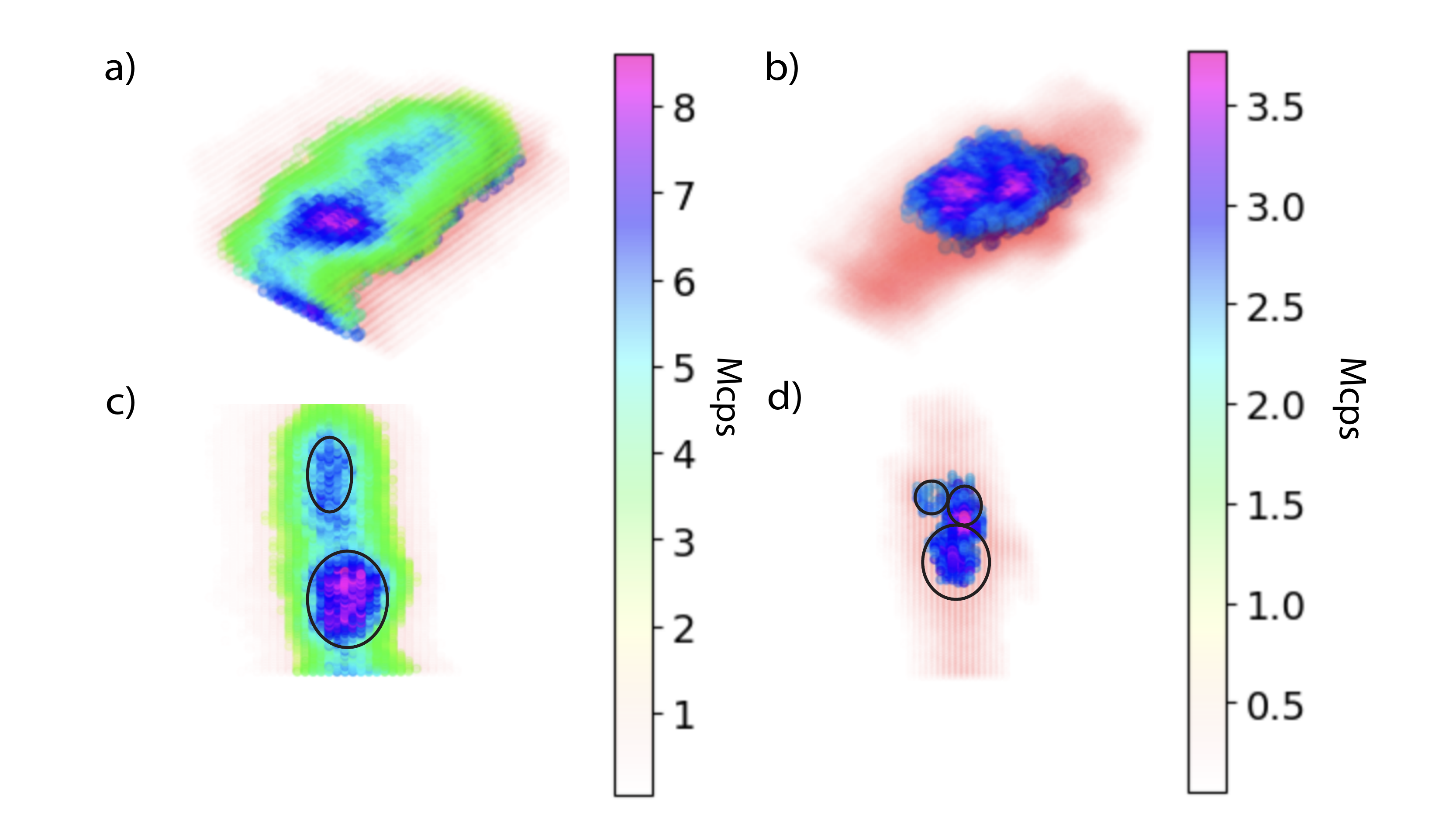}
\caption{Focus-dependent 3D PL Maps. Opaque blue coloring corresponds to high counts, while transparent red corresponds to low counts.
(a) PL map obtained in the C/C configuration. Steps in the focal plane are reflected in the plot by different z-axis locations. PL maps in 2D are stacked along the z-axis to produce a 3D image. We threshold the data to make lower count areas more transparent for visualization purposes.
(b) PL map obtained in the C/CP configuration. Configuration of data and thresholding are identical to that used to produce the map in (a).
(c) Top-down view of two focal plane steps from (a). We circle the suspected clusters of emitters for visualization purposes.
(d) Top-down view of two focal plane steps from (b) (same focal steps as in (c)). We circle the suspected clusters of emitters for visualization purposes.
} 
\label{fig:supfig3c}
\end{figure}

We also collect full PL maps in the X-Y plane at each interrogated focal plane. For each collection configuration, we overlay this data, with the $z$ axis corresponding to focal plane shifts, thereby generating a 3D map of suspected emitter locations. The results, shown in Fig.~\ref{fig:supfig3c}, further accentuate the differences between the two collection configurations. First, by comparing Fig.~\ref{fig:supfig3c}(a) and Fig.~\ref{fig:supfig3c}(b), we observe greater background signal and increased blurring of predicted emitter locations in the C/C configuration (Fig.~\ref{fig:supfig3c}(c)). However, in the C/CP configuration, we more clearly identify what we believe to be small clusters of emitters (Figure~\ref{fig:supfig3c}(d)). Overall, the data suggests that the TFLN waveguide collection channels are less sensitive to background fluorescence as compared to the confocal collection channels. This difference might be useful in future studies for resolving emitters in 3D space. However, the waveguide collection channels may be more susceptible to collecting in-nanobeam scattering and coupling to un-targeted emitters. Furthermore, our collection here must be calibrated to the in/out-coupling of the gratings, as described in sections 4 and 7.

\end{document}